\documentclass[twocolumn,aps,floatfix,showpacs,prb,superscriptaddress]{revtex4}
\usepackage{epsfig}
\usepackage{graphicx}
\usepackage{amsmath}
\usepackage{amssymb}
\usepackage{amsfonts}
\usepackage{tabularx}
\usepackage{color}
\usepackage{dcolumn}% Align table columns on decimal point
\usepackage{bm}% bold math
\usepackage{floatflt}
\usepackage{setspace}

\newcommand{\blue}{\color{blue}}

\newcommand{\urusi}  {URu$_2$Si$_2$}

\begin{document}

\author{P. M. Oppeneer}
\affiliation{Department of  Physics and Astronomy, Uppsala University, Box 516, S-75120 Uppsala, Sweden}
\author{J. Rusz}
\affiliation{Department of  Physics and Astronomy, Uppsala University, Box 516, S-75120 Uppsala, Sweden}
%\altaffiliation[also at ]{Institute of Physics, Academy of Sciences of the Czech Republic, Na Slovance 2, CZ-182 21 Prague, Czech Republic.}
\author{S. Elgazzar}
\altaffiliation[On ]{ leave from Menoufia University, Menoufia, Egypt.}
\affiliation{Department of Physics and Astronomy, Uppsala University, Box 516, S-75120 Uppsala, Sweden}
\author{M.-T. Suzuki}
\affiliation{Department of Physics and Astronomy, Uppsala University, Box 516, S-75120 Uppsala, Sweden}
\author{T. Durakiewicz}
\affiliation{Los Alamos National Laboratory, Condensed Matter and Thermal Physics Group, Los Alamos, NM 87545,USA}
\author{J. A. Mydosh}
\affiliation{Kamerlingh Onnes Laboratory, Leiden University, NL-2300 RA
Leiden, The Netherlands}

\title{Electronic structure theory of the hidden order material {\urusi}}

\date{\today}

\begin{abstract}
We report a comprehensive electronic structure investigation of the paramagnetic (PM), the large moment antiferromagnetic (LMAF), and the hidden order (HO) phases of {\urusi}. We have performed relativistic full-potential calculations on the basis of the density functional theory (DFT),
employing different exchange-correlation functionals to treat electron correlations within the open $5f$-shell of uranium. Specifically, we investigate---through a comparison between calculated and low-temperature experimental properties---whether the $5f$ electrons are localized or delocalized in {\urusi}.   The local spin-density approximation (LSDA) and generalized gradient approximation (GGA) are adopted to explore itinerant $5f$ behavior,  the GGA plus additional strong Coulomb interaction (GGA+$U$ approach) is used to approximate moderately localized $5f$ states, and the $5f$-core approximation is applied to study completely localized uranium $5f$ states.
We also performed dynamical mean field theory calculations (LDA+DMFT) to investigate the temperature evolution of the quasi-particle states at 100~K and above, unveiling a progressive opening of a quasi-particle gap at the chemical potential when temperature is reduced.
A detailed comparison of calculated properties with known experimental data demonstrates that the LSDA and GGA approaches, in which the uranium $5f$ electrons are treated as itinerant, provide an excellent explanation of the available low-temperature experimental data of the PM and LMAF phases. 
%The investigated quantities including the equilibrium volume, internal coordinates, bulk modulus, the spin and orbital magnetic moment, the Fermi surface gapping, number of carriers, resistivity anisotropy and resistivity change, de Haas-van Alphen and Shubnikov-de Haas quantum oscillations, and Fermi surface nesting vectors. 
%
We show furthermore that due to a materials-specific Fermi surface instability a large, but partial, Fermi surface gapping of up to 750 K occurs upon antiferromagnetic symmetry breaking. 
The occurrence of the HO phase is explained through dynamical symmetry breaking induced by a mode of long-lived antiferromagnetic spin-fluctuations. This dynamical symmetry breaking model  explains why the Fermi surface gapping in the HO phase is similar but smaller than that in the LMAF phase and it also explains why the HO and LMAF phases have the same Fermi surfaces yet different order parameters. Suitable derived order parameters for the HO are proposed to be the Fermi surface gap or the dynamic spin-spin correlation function.
%involving both time-reversal symmetry breaking and lattice symmetry breaking. 
%Temperature-dependent  DMFT calculations reveal the progressive opening of a quasi-particle gap when temperature is reduced.  
\end{abstract}

%\keywords{hidden order, {\urusi}, dynamical symmetry breaking, longitudinal spin-fluctuations, density functional theory, dynamical mean field theory}

\maketitle

\section{Introduction}

Over the past 15 years the concept of ``hidden order" has evolved to 
describe the emergent behavior of various quantum or strongly correlated 
materials where the order parameter of a clear phase transition along with 
its elementary excitations remain unknown. Often modern microscopic 
measurement techniques of diffraction (neutrons or photons), nuclear magnetic resonance (NMR), or muon spin-rotation  ($\mu$SR), 
etc. are unable to detect and characterize the new ordered phase. Yet the 
thermodynamic and transport properties unambiguously discern a novel 
state of matter appearing at a sharp transition temperature. Within this 
state additional unconventional phases may form depending upon varying 
parameters such as pressure, magnetic (electric) fields, and doping.  
Although there is at present no comprehensive review of the generic 
hidden order (HO) problem and its relation to quantum criticality, the HO
concept is beginning to make headway into the recent literature. 
\cite{shah00,coleman05,dallatorre06,xu07,hassinger08b,santini09}

A prototype system for this behavior is the intermetallic compound 
{\urusi}, discovered 25 years ago.\cite{palstra85, schlabitz86,maple86}
This material displays strong 
electron correlations such that the U 5$f$  magnetic moments are dissolved 
into hybridized bands near the Fermi surface and a moderately heavy Fermi liquid 
forms at temperatures below ca.\ 70~K.\cite{schoenes87,pfleiderer06} 
%WHAT ABOUT TK?
Then at 17.5~K the HO state appears 
via a dramatic (second-order) phase transition.\cite{palstra85,schlabitz86,maple86}
 The above-mentioned 
techniques fail to discern the order parameter and cannot characterize its 
elementary excitations. Great attention has been devoted to studying this 
system with the aim of uncovering its hidden nature. A vast collection of 
experimental data has been gained and excellent single crystals are now 
available for definitive investigations thereby eliminating extrinsic effects 
of  impurities and stress (see, e.g., Ref.\ \onlinecite{amitsuka07}).
In addition there are numerous theoretical 
proposals and exotic models spanning many years\cite{barzykin95,santini94,kasuya97,okuno98,ikeda98,chandra02,mineev05,kiss05,hanzawa05,varma06,hanzawa07,elgazzar09,balatsky09,cricchio09,haule09,harima10} that have, however, not come to 
full grips with many aspects of the experimental behavior.

Recent investigations\cite{amato04,amitsuka07,jeffries07,hassinger08,aoki09,niklowitz10} on good single crystals 
 have mapped out the phase diagram of {\urusi}.
Apart from the paramagnetic (PM)  phase and the HO phase below 17.5 K at ambient pressure,
there is also the large moment antiferromagnetic (LMAF) phase, which appears with modest pressure of about 0.5 GPa and is characterized by uranium moments of 0.4 $\mu_B$ in a type-I antiferromagnetic (AF) arrangement. Surprisingly, 
the bulk properties of the HO and LMAF phases are very much alike. Very similar, continuous changes in the thermodynamic and transport quantities have been reported for both phases.\cite{mcelfresh87,amitsuka07,jeffries07,hassinger08,bourdarot10} A comparable Fermi surface gapping occurs for the transitions from the PM phase to the HO and LMAF phases, respectively. This similarity-- which has been called adiabatic continuity\cite{jo07}-- extends to the Fermi surfaces of the HO and LMAF phases. 
De Haas-van Alphen experiments detect no significant differences between the Fermi surfaces of the HO and LMAF phases\cite{nakashima03}  and, consistently, neutron scattering experiments find the same nesting vectors.\cite{villaume08} Nonetheless, the HO and LMAF unmistakably have different order parameters; simple magnetic order in the LMAF phase but an unknown order parameter in the HO phase. Neutron and x-ray scattering experiments\cite{broholm87,mason90,isaacs90,broholm91,walker93}
 detected a small magnetic moment $\sim$0.03~$\mu_B$ in the HO phase, but this small moment is currently considered as a parasitic moment that is not intrinsic to the HO phase.\cite{amitsuka07,takagi07,niklowitz10} 
 % WHERE
% The HO transition is accompanied by a large entropy reduction\cite{maple86} which cannot be explained by small static moments.\cite{buyers96}
Other differences between the HO and LMAF phases is that 
%nowadays   a vanishing small versus  order LMAF phase is a  as the   [8-10]. 
below 1.2~K and only out of the HO an unconventional\cite{matsuda96} superconducting state appears, which is the subject of recent interest.\cite{kasahara07,yano08,kasahara09} A further salient difference between the HO and LMAF phases is that inelastic neutron experiments detected a mode of AF spin-fluctuations in the HO phase which freezes to the static antiferromagnetic Bragg peak in the LMAF phase.\cite{broholm91,villaume08,bourdarot10}

As a starting point towards a full theoretical understanding of the intriguing electronic structure of {\urusi} state-of-the-art band structure calculations are required. We present here detailed investigations of the electronic structures of the PM and LMAF phases, using various computational methods. On the basis of the obtained electronic structures, we  analyze in how far the known physical properties of {\urusi} can be explained from these underlying electronic structures, and draw conclusions on what the valid electronic structure of {\urusi} is, emerging from the electronic structure calculations. Subsequently, we focus on the implications for a prospective explanation of the HO.
Also, we expand on the ``dynamical symmetry breaking" model for the HO, which we have recently proposed.\cite{elgazzar09} Details of this model are given and we relate the model to a larger collection of experimental properties. We also compare the derived electronic structure and the HO model to other recent proposals. 

In the following we first consider an issue that is central to the current discussion of model explanations of the HO phase.

%equilibrium volume, lattice coordinates, equation of state, 
%nesting vectors, de Haas-van Alphen quantum oscillations,
%Fermi surface gapping, resistivity, number of carriers, compensated metal, 
% magnetic spin and orbital moment

\section{Itinerant or localized $5f$ behavior?}

One of the most intriguing questions regarding the electronic structure  of {\urusi}, and consequently the explanation of the HO, is whether the uranium $5f$'s are localized or delocalized. Single-ion theories of the HO, such as, e.g., quadrupolar or octupolar ordering, are based on the assumption of localized $5f$'s.
\cite{santini94,sikkema96,okuno98,ohkawa99,kiss05,fazekas05,hanzawa05,hanzawa07,haule09,harima10,haule10} 
This important issue of the degree of $5f$ localization has been controversially discussed recently.  
Several theories adopt the picture of localized $5f$ states from the outset, however, an examination of the grounds for this is needed. 
A thorough examination seems to unveil that there is little compelling experimental evidence for localized $5f$'s. Smoking-gun evidence for localized $f$ states would be the classical observation of crystal electrical field (CEF) $f$ excitations in neutron experiments, but only itinerant spin excitations have been detected and CEF excitations have never been observed for {\urusi} (see, e.g., Ref.\ \onlinecite{wiebe07}). Another indication of a CEF excitation could, e.g., come from measured specific heat curves, in which humps or peaks could signal the occurrence of CEF excitations. The measured $C/T$ curve of {\urusi} shows a maximum at 70 K,\cite{schlabitz86} which has sometimes been interpreted as evidence for a CEF transition. However, later measurements\cite{janik08} of the $C/T$ of ThRu$_2$Si$_2$, which has no occupied $5f$'s and hence no CEFs, revealed  a very similar maximum at the very same temperature. This suggests that the peak at 70 K is more likely related to the same underlying lattice
%phonon 
structure and not to a CEF transition of $5f$ states. 
The shape of the measured magnetic entropy $S_m (T)$ in the PM state does not correspond to Schottky-type anomaly expected for CEF levels.\cite{vandijk97,janik08}
Also, very recent scanning tunneling spectroscopy (STS) measurements could not detect any CEF splitting of the $5f$'s.\cite{aynajian10}
Consistently, the susceptibility of {\urusi} does not show Curie-Weiss behavior near the HO temperature that might indicate localized $f$ states, rather Curie-Weiss behavior commences only above 150~K.\cite{palstra85,yokoyama02}

One particular piece of experimental evidence in favor of localized $5f$'s  has come from inelastic neutron scattering experiments,\cite{park02} in which a small inelastic peak was observed at 363 meV. This peak has been interpreted as a signature of an inter-multiplet transition.\cite{park02} A similar peak has been observed for UO$_2$, which is indeed known to have a localized $5f^2$ configuration. However, for UO$_2$ CEF excitations, too, were definitely observed with inelastic neutron scattering (see, e.g., Ref.\ \onlinecite{holland94}).
 Inelastic neutron experiments\cite{park02} also detected a small peak at 363 meV for ThRu$_2$Si$_2$, which indicates that the peak might not be due to an inter-multiplet excitation. In addition, a similar 
 %inter-multiplet 
 peak has been observed\cite{hiess97} for URhAl, which is however known to be an itinerant $5f$ material.\cite{paixao93} 
The origin of the neutron peak at 380 meV in URhAl has consequently been debated;\cite{kunes01b} the issue is not completely solved, but it could be an artifact related to the measurement apparatus.

Several other experimental data rather advocate the existence of delocalized $5f$ electrons in {\urusi}.
High-resolution photoemission spectroscopy (PES) using He~I and He~II radiation gave evidence for a typical delocalized $5f$ response in the He~II$-$He~I difference spectrum.\cite{yang96}
A similar difference spectrum has been observed for itinerant U-metal and UGa$_3$.\cite{gouder01}
In addition, angular resolved photoemission spectroscopy (ARPES) revealed dispersive bands in {\urusi},\cite{ito99,denlinger01} yet it still needs to be clarified what the dominant character of the observed  bands is ($f$ related or not).  On the other hand, very recent He~I ARPES measurements provided a picture of an almost flat band which sinks through $E_F$ at the HO transition.\cite{santander09} The picture of a  narrow band very close to $E_F$ may however arise from the special data treatment, i.e. division by the Fermi function and double-differentiation technique) which always tends to give an impression of a flat state near $E_F$.%\cite{durakiewicz09} 

The AF phase of {\urusi} is commonly referred to as the large moment antiferromagnetic phase. This name suggests that the $5f$'s in the LMAF phase might be partially localized.
 However, in spite of its name, the uranium moment in the LMAF phase is actually relatively small and not the typical moment of a localized $5f$ material. For example, the $5f$ states of the cubic uranium salt USe are known to be closer to $5f$ localization, but still exhibit some $f-d$ hybridization, which leads typically to spin, orbital, and total moments of  -1.1, 3.1, and 2.0 $\mu_B$, respectively, for USe.\cite{hashimoto98} The total moment on U in AF {\urusi} is with 0.4 $\mu_B$ quite far from such value. Instead, the U moment is much closer to values of 0.6 $\mu_B$ measured for an itinerant $5f$ material such as UGa$_3$.\cite{nakamura02}

It also deserves to be mentioned that the results of recent positron annihilation experiments on {\urusi} proved to be incompatible with localized $f$'s, but are in contrary in good agreement with delocalized $f$'s.\cite{biasini09} Also, recent neutron scattering experiments detected itinerant $5f$ spin excitations.\cite{wiebe07}

Altogether, there does not appear to be clear, compelling evidence for localization of $5f$ electrons in {\urusi}. On the other hand, there exists a body of evidence in favor of delocalized $5f$'s. Nonetheless, the decision on localized-itinerant behavior should be concluded from an extensive comparison of calculated and experimental properties, which will be presented below.

 In the following Section we first outline the here-to-be applied first-principles based techniques to study the electronic structure of {\urusi}. With these different approaches we can treat the full range of $5f$ behavior, from delocalized to localized. In view of the above considerations regarding the itinerant or localized $5f$ behavior, our main focus will be on the $5f$ band description. The applied DFT-LSDA and LSDA+$U$ approaches can provide only ground-state $T=0$~K properties. The temperature dependence of quasi-particle spectra will be treated through dynamical mean field theory (DMFT) calculations.

\section{Computational Methodology}

Our calculations are based on the DFT as well as on the DMFT. Specifically, for the treatment of the DFT static exchange-correlation potential we have employed the local spin density approximation (LSDA),
% in the Perdew-Wang parametrization,
\cite{perdew92} the generalized gradient approximation (GGA),\cite{perdew96} and also orbital-dependent extensions (LSDA+$U$, GGA+$U$) to include the influence of strong on-site Coulomb correlations.

In our calculations we have used three accurate full-potential, relativistic electronic structure codes. These are the full-potential local orbitals (FPLO) method\cite{koepernik99,eschrig04} and the full-potential linearized augmented plane wave (FLAPW) method; the latter we employed both in the 
 WIEN2K\cite{wien2k} and Kansai implementations.
 We have verified that the three codes give, on the self-consistent LSDA level, {\it identical} results for the electronic structure of {\urusi}.

In the  FLAPW calculations the relativistic valence states are computed within the full, non-spherical potential. The relativistic spin-orbit interaction (SO) was included self-consistently,\cite{kunes01b} and, in the WIEN2K calculations, we used the relativistic local orbitals extension of the scalar-relativistic FLAPW  basis to treat accurately the $2p_{1/2}$ semi-core states.\cite{kunes01a}
 The product of $R_{mt}$ and maximum reciprocal space vector ($K_{max}$), i.e., the basis size determining parameter ($RK_{max}$) was set to 7.5 and the largest reciprocal vector $\boldsymbol{G}$ in the charge Fourier expansion, $G_{max}$,  was equal to 12. We used about 5000 $k$-points for self-consistent convergence.
With the  WIEN2K calculations we have employed the orbital-dependent  GGA+$U$ method with around mean-field double-counting correction, \cite{czyzyk94} in which  an additional on-site Coulomb interaction, expressed by the Hubbard $U$ and exchange $J$ parameters, is introduced for the $5f$-states manifold. 
%{\blue The values chosen for $U$ and $J$ will be discussed below. -needed?}

In the relativistic full-potential FPLO calculations\cite{eschrig04}
%\cite{opahle01,eschrig04}
the four-component Kohn-Sham-Dirac equation, which implicitly
contains spin-orbit coupling up to all orders, is solved self-consistently. 
We  used in the FPLO calculations the following sets of basis orbitals: $5f$; $6s6p6d$; $7s7p$ for U, $4s4p4d$; $5s5p$, and $3s3p3d$, for Ru and Si, respectively.
 The high-lying $6s$ and $6p$ U semicore states, which might hybridize with other valence states  are thus included in the basis. The site-centered potentials and densities were expanded 
in spherical harmonic contributions up to $l_{max}=12$. 
Brillouin zone (BZ) sampling was performed with maximally 20$\times$20$\times$20  $\boldsymbol{k}$-points. 

For the DMFT calculations we have used a recently developed\cite{suzuki09} 
%fully-self-consistent, 
full-potential, relativistic LSDA+DMFT method. For a  detailed review of the  DMFT method we refer to Ref.\ \onlinecite{kotliar06}. 
In our DMFT calculations we use
the spin-polarized T-matrix fluctuation-exchange (FLEX) impurity solver \cite{bickers89,lichtenstein98}  for generating the self-energy. This impurity solver is expected to be applicable to moderately correlated materials, as, e.g., uranium intermetallic compounds.
The local Green function is computed 
employing Kohn-Sham states which are obtained from a relativistic LSDA+$U$ calculation.
The Coulomb $U$ and exchange $J$ quantities are connected to the two-electron integrals of the Coulomb interaction of the $f$ electrons through the  effective Slater integrals $F_{\kappa}$ ($0 \le \kappa \le 6$),  where $F_0=U$,  $F_4=\frac{41}{297}F_2$, $F_6=\frac{175}{11583}F_2$, and $J=\frac{286F_2 + 195F_4 + 250F_6}{6435}$.
Overall self-consistency  has been achieved through iterative feeding of the density matrix  of the local Green function in the next LSDA+$U$ loop, and back feeding of the new solutions in the DMFT loop.
In the DMFT calculation we used 8192 Matsubara frequency points to compute the temperature dependent quasi-particle spectrum, which was obtained using a Pad{\'e} approximation to the frequency dependent lattice Green function $G ({\mbox{\boldmath$k$}},i \omega) $. Within the present DMFT FLEX implementation temperatures down to about 100 K can only be be reached.

\begin{figure}[tbh]
   \includegraphics[width=0.24\textwidth]{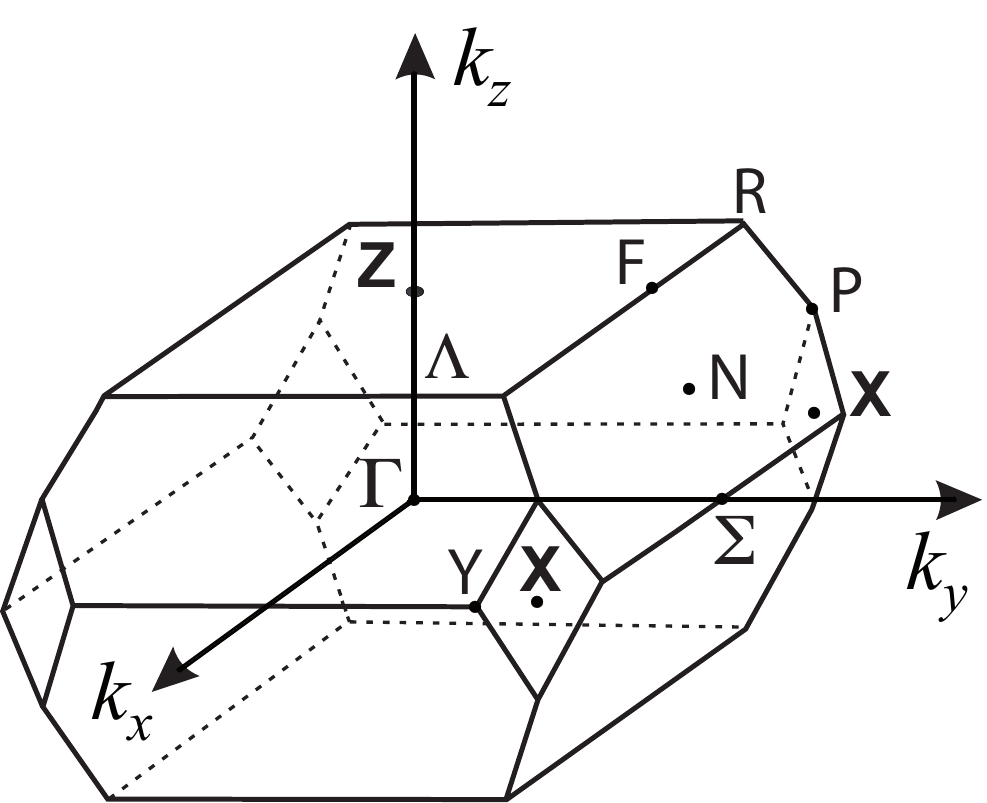}
  \includegraphics[width=0.23\textwidth]{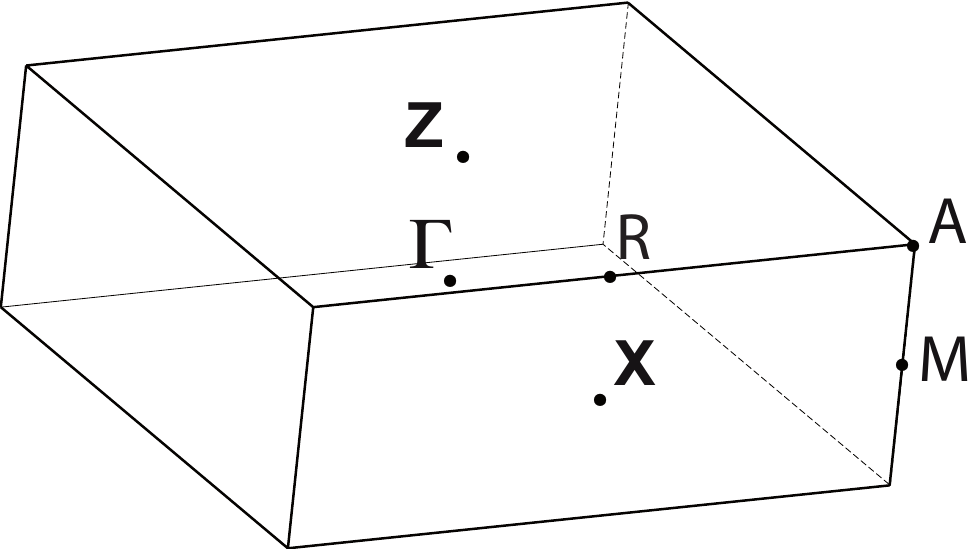}
  \caption{The Brillouin zones of the body centered  tetragonal phase (space group No.\ 139, left)  and the simple tetragonal phase (space group No.\ 123, right) with high-symmetry points indicated.\cite{note1}
  \label{fig:BZ}}
\end{figure}

In our investigations we employ the body-centered tetragonal  (i.e., ThCr$_2$Si$_2$) structure with space group No.\ 139 for paramagnetic {\urusi} and the simple tetragonal structure with space group No.\ 123 for AF {\urusi}. The \textsc{st} unit cell volume of AF {\urusi} is twice the \textsc{bct} unit cell volume of PM {\urusi}. The space group of the HO phase has not yet been definitely established, because the symmetry breaking in the HO phase is as yet to be unveiled. 
In Fig.\ \ref{fig:BZ} the Brillouin zones of the \textsc{bct} and \textsc{st} structures are shown with high-symmetry points indicated. The \textsc{st} BZ corresponds to a folding of the \textsc{bct} BZ at $\pm \frac{1}{2}$Z. 
In the \textsc{bct} BZ we have additionally labeled several non-high-symmetry points ($\Sigma$, F, Y, and $\Lambda$) for later discussion.

\section{Results}

\subsection{DFT delocalized $5f$ electron calculations}

\subsubsection{Structural optimization}

\begin{figure}[tbh] 
 \includegraphics[width=0.45\textwidth]{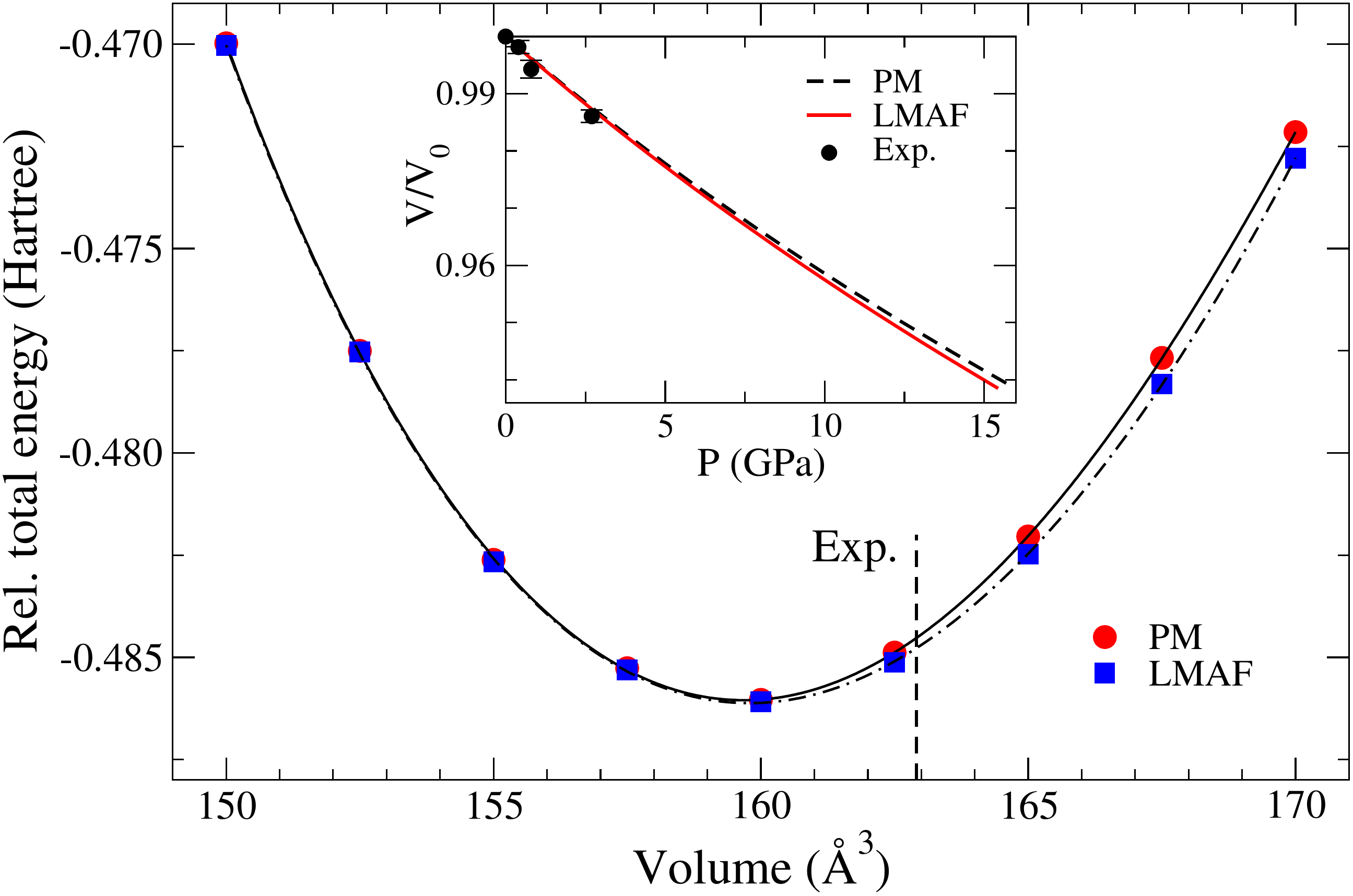}
  \caption{(Color online)  The {\it ab initio} computed total energy versus volume for the PM and LMAF phases of {\urusi}. 
  The vertical dashed line indicates the experimental equilibrium volume,\cite{cordier85}
  the inset shows the {\it ab initio} computed pressure dependence of the unit cell volume for the PM and LMAF phases, normalized to the zero-pressure volume $V_0$, together with experimental data points of Ref.\ \onlinecite{jeffries-unp}. 
  \label{fig:volume}}
\end{figure}

To start with, we consider the structural properties of {\urusi}, that is, the equilibrium lattice coordinates, 
bulk modulus, and equation of state.
Several experimental investigations of the structural properties of {\urusi} have been reported.\cite{palstra85,jeffries-unp,cordier85,motoyama03}
For comparison to the available data, we have performed {\it ab initio} optimizations of the equilibrium volume, the $c/a$ ratio, and the internal Si coordinate, $z_{\rm Si}$. These optimizations have been performed on the LSDA level, both for the PM and LMAF phases. 
In Fig.\ \ref{fig:volume} we show the computed total energy versus unit cell volume. Both PM and LMAF total energies are given for the double unit cell, to convene comparing the two phases. The theoretically predicted equilibrium unit cell volume is about 1.7\% smaller than the experimental volumes, which are 162.9 {\AA}$^3$ (Refs.\ \onlinecite{palstra85} and \onlinecite{cordier85}), respectively, 162.6 {\AA}$^3$ (Ref.\ \onlinecite{jeffries-unp}). Hence, the theoretical value is in very good agreement with experiment. As Fig.\ \ref{fig:volume} illustrates the total energies of the PM and LMAF phases are very near one another. The total energy of the LMAF phase is computed to be only 7 K per formula unit deeper than that of the PM phase. This is in itself a remarkable finding, which appears to be a specific feature of {\urusi}. The inset of Fig.\ \ref{fig:volume} presents the computed volume versus pressure dependence of {\urusi}. With pressure the antiferromagnetic state becomes slightly more stable. The inset includes recent experimental data points of Ref.\ \onlinecite{jeffries-unp}. A fit of the computed volume versus pressure curves gives a bulk modulus $B_0$ of 204 GPa (208 GPa) for the LMAF (PM) phase. The recent pressure experiment\cite{jeffries-unp} obtained a value of 190 GPa; an older experiment reported a value of 230 GPa.\cite{luo88}

\begin{figure}[tbh]
  \includegraphics[width=0.4\textwidth]{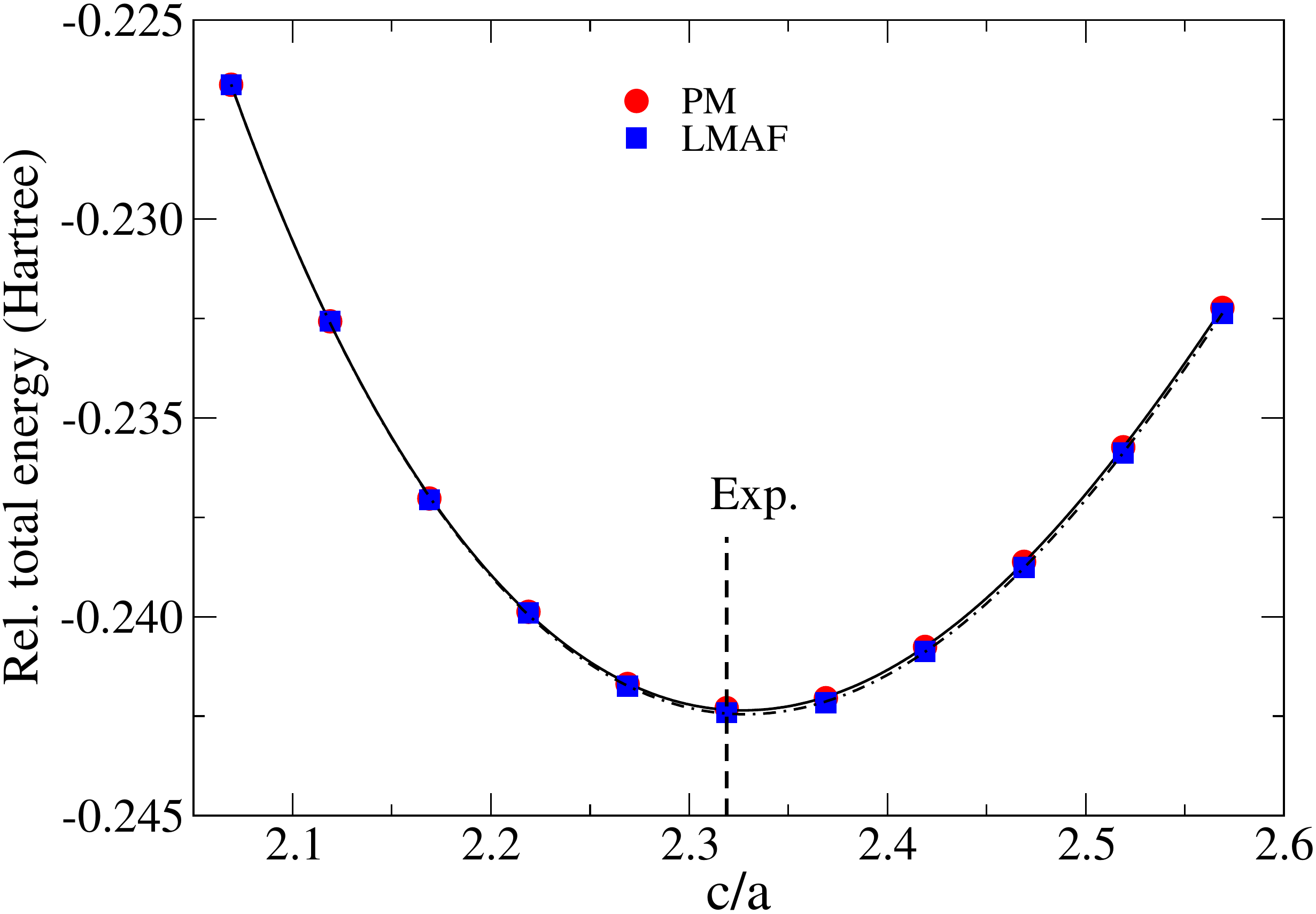}
  \caption{(Color online)  Total energy optimization of the $c/a$ ratio of {\urusi} for the
  PM and LMAF phases. The dashed vertical line denotes the experimental value.\cite{palstra85,cordier85}
   \label{fig:coa}}
\end{figure}

The optimized theoretical $c/a$ ratio is shown in Fig.\ \ref{fig:coa}. The obtained theoretical $c/a$ ratio almost coincides with the experimental value (2.32).\cite{palstra85,cordier85}
The optimized $c/a$ ratio of the LMAF phase is found to be just a small fraction larger than that of the PM phase. We note that x-ray diffraction experiments have been unable to detect any difference in the lattice constants of the  PM, LMAF, and HO phases.\cite{jeffries-unp,kuwahara03}
Only dilatation experiments\cite{motoyama03,niklowitz10} could so far detect tiny differences in both the $a$ and $c$ lattice constants of the three phases; the $c$ axis lattice constant of the LMAF and HO phases are elongated with a few parts in  $10^{-5}$, as compared to the PM phase (at higher temperature). The $a$ axis of the HO and LMAF phases is contracted by a few parts in $10^{-5}$. As a result, the $c/a$ ratio increases\cite{motoyama03} with about $10^{-4}$ from the PM above 17.5 K down to the LMAF phase at 10 K.
The optimized $c/a$ ratio of the LMAF phase is consistently computed here to be about $10^{-4}$ larger than that of the PM phase, in agreement with experiment.\cite{motoyama03} We refrain
however from a more detailed comparison because we cannot make a meaningful quantitative statement for such tiny numbers.
 
Using the optimized $c$ axis lattice constant and volume, the theoretical $a$ axis lattice constant is about 0.6\% smaller than the experimental value.\cite{palstra85,cordier85,jeffries-unp} As the LSDA approach is know to produce a small overbinding, the correspondence with the experimental lattice constant can be regarded as very good.

\begin{figure}[tbh]
  \includegraphics[width=0.4\textwidth]{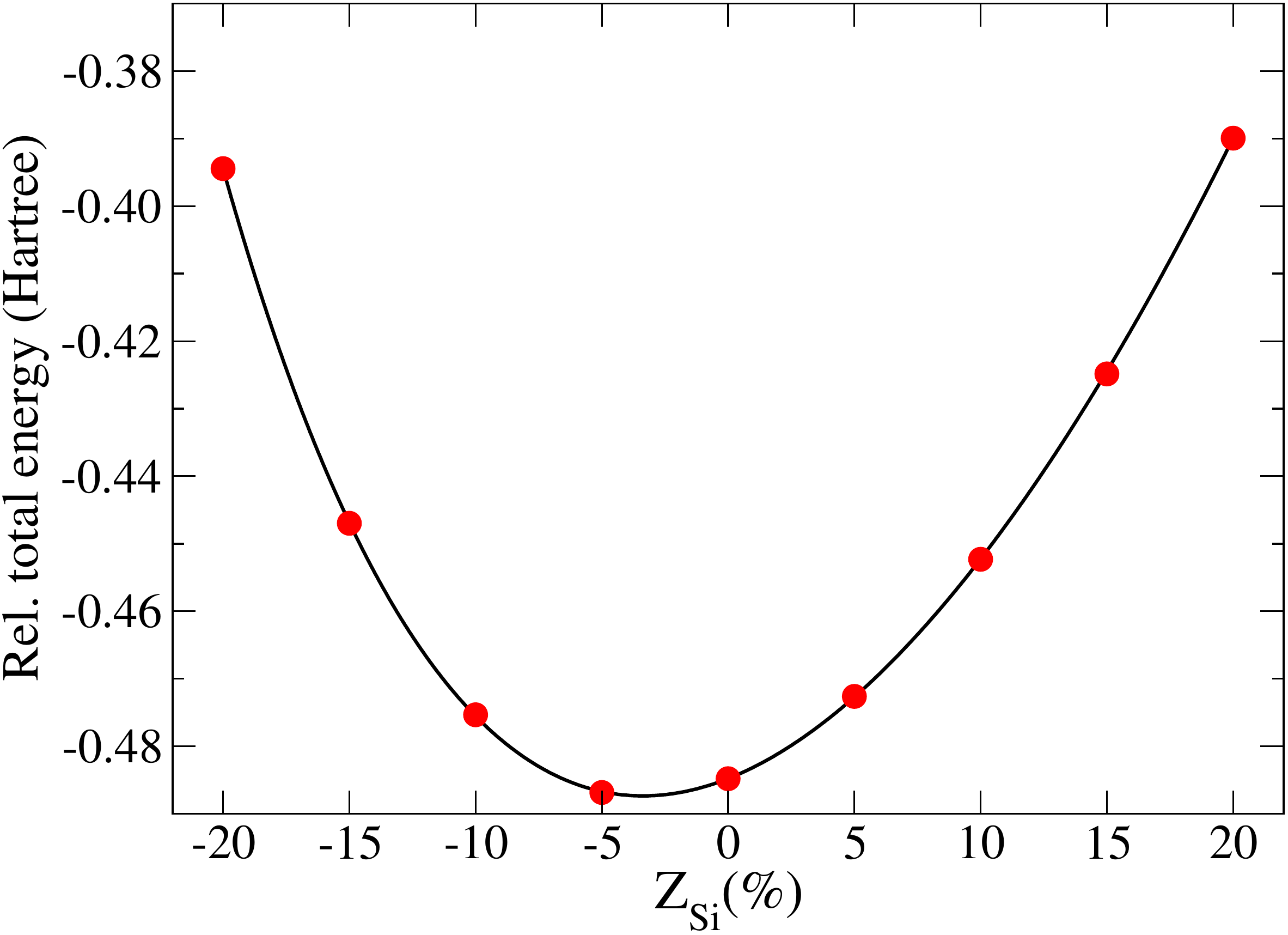}
  \caption{(Color online)  Total energy optimization of the special Si $z$-position in {\urusi}. 
  The zero position of $z_{\rm Si}$ is taken at 0.371 (Ref.\ \onlinecite{cordier85}).
   \label{fig:Si_z}}
\end{figure}

The \textsc{bct} structure of {\urusi} has one internal coordinate, the Si $z$-position. 
Results for the total-energy optimization of the $z_{\rm Si}$ coordinate are given in Fig.\ \ref{fig:Si_z}.
The theoretical LSDA value is found to be 3\% smaller than the experimental value of 0.371 (Ref.\ \onlinecite{cordier85}). A more recent experiment\cite{jeffries-unp} obtained a somewhat smaller $z_{\rm Si}$ value, 0.3609, which would agree quite well with the theoretical result.
As will be shown in more detail below, the essential physical properties of {\urusi} are stable with respect to moderate variations of the unit cell dimensions and the internal $z_{\rm Si}$ coordinate.

Altogether, the {\it ab initio} structural optimization shows that the crystallographic properties of {\urusi} are well described by the LSDA approach, which intrinsically  is based on the assumption of itinerant $5f$ electrons. It is known from computational investigations for other actinide materials that, when these have $5f$ states that are localized, the LSDA approach usually does not provide a good description of the lattice properties (see, e.g., Ref.\ \onlinecite{shick06}). Such deviant behavior is not found here for {\urusi}.

%Values for optimization:
%exp. a=4.1262, c=4.5684, c/a=2.3189, ZSi=0.371 Vol=162.91 (Cordier, check)
%Palstra a=4.1239, c=9.5817, c/a=2.2334

%Jeffries: zsi=0.3609, B0=190, a=4.1252, c=9.5578 Vol=162.648

%Our calculations give: zSi-3 proc= 0.360, c/a=2.32(5)
%Vol is 1.78 proc. smaller, lattice constant 0.6 proc. or 0.57 proc.
%B0 is 204 (208)
%Vol. is 1.63 proc smaller for Jeffries measurement

%Moments are stable versus zSi position.

%EOS: good agreement with Jeffries. AFM more stable under pressure.
%Total energy difference: AFM deeper by 7K f.u.

%c over a slightly larger for AFM phase, as found in experiments (exps. give 1.2 10-4, but calcs about 10 10-4). However, this is difficult with error bars.
%X-ray diffraction experiments cannot detect crystallographic differences between the phases.

%Motoyama et al: Delta a/a= -4 10-5, Delta c/c < 1.10-5 PM/HO
%Delta a/a= -7 10-5, Delta c/c = 5.10-5 PM/LMAF

%\subsection{Equilibrium lattice quantities}

\subsubsection{Energy band dispersions}

\begin{figure}[thb]
  \includegraphics[width=0.45\textwidth]{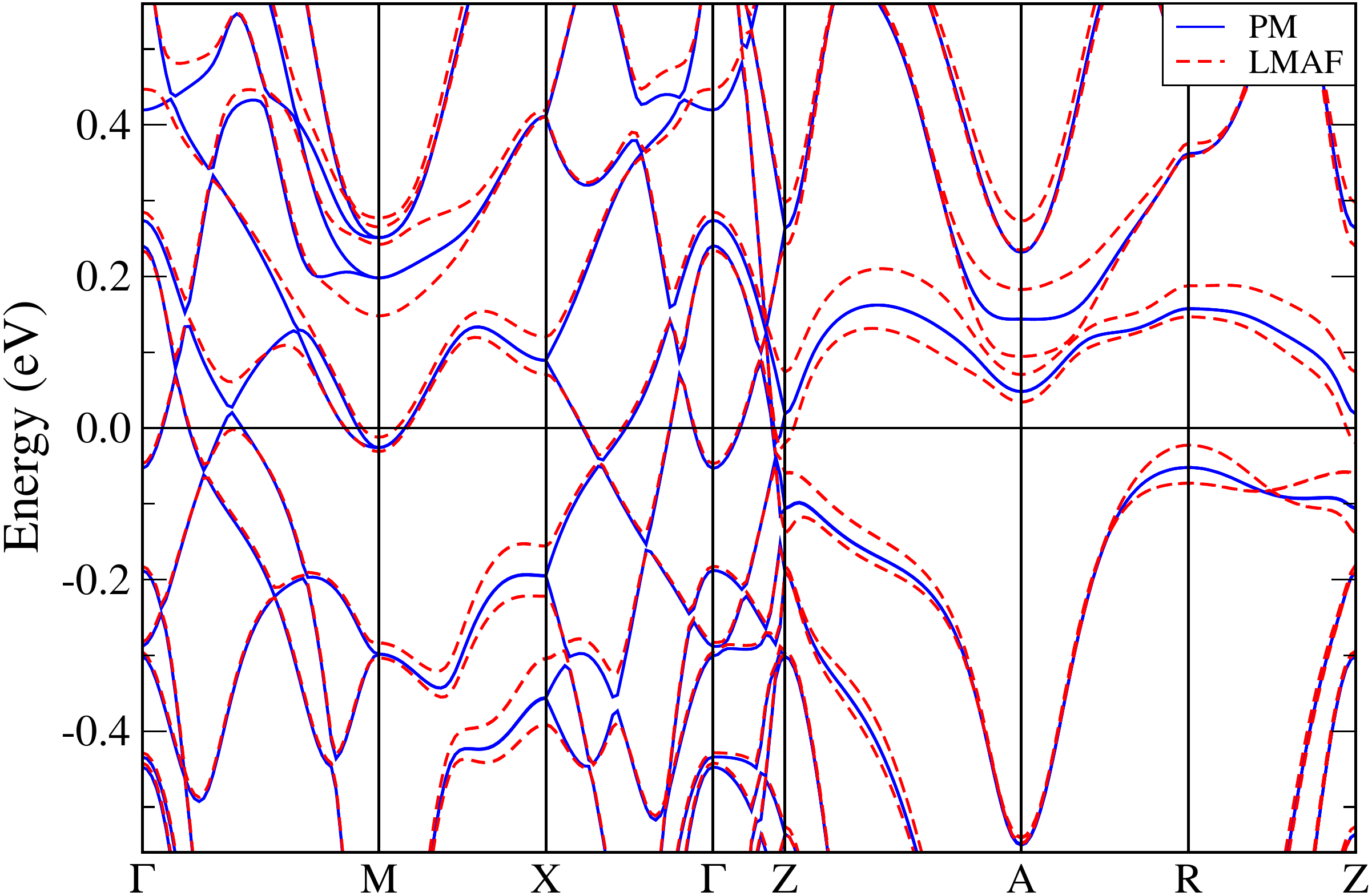}
  \caption{(Color online)  Computed LSDA energy dispersions of {\urusi} in the PM and LMAF phases, both are shown, for sake of comparison, in the simple tetragonal BZ. 
   \label{fig:bands-123}}
\end{figure}

We first consider the outcome of LSDA $5f$-itinerant  calculations for the electronic structure of {\urusi}.
In Fig.\ \ref{fig:bands-123} we show the computed LSDA energy dispersions in the PM and LMAF phases for the experimental lattice parameters.\cite{cordier85} %Again, 
To draw a comparison, both sets of dispersions are given for the double unit cell (space group No.\ 123). As has been noted recently by us, the energy dispersions of these two phases are very similar.\cite{elgazzar09} The dispersions of the AF phase are almost on top of those of the PM phase, except for some influence of the exchange splitting of $5f$ related bands. This finding corroborates fully with the compute tiny total energy difference between these two phases.
A degeneracy of crossing bands occurs near the Fermi level, as can be recognized along the $\Gamma -$M and X$-\Gamma$ symmetry directions. The degenerate band crossing, existing  in the PM phase along the $\Gamma -$M direction just below $E_F$, is lifted in the LMAF phase, due to a re-hybridization, and thereby a small gap opens. A similar degenerate crossing point along the X$- \Gamma$ direction is however  not removed in the  LMAF phase. Through a larger part of the BZ degenerate crossings of the two bands exist, yet the opening of a gap in the AF phase does not happen uniformly over the Fermi surface (FS).\cite{elgazzar09} This gapping is related to a FS instability of {\urusi} in the PM phase, where degenerate band crossing (``Dirac points") occur  off the high-symmetry directions, between the $\Gamma - $M and $\Gamma -$X directions. These degenerate points are removed in a transition to the LMAF phase, leading to a $k$-dependent FS gapping that is largest in the $z=0$ plane.\cite{elgazzar09}

 \begin{figure}[tb]
 \begin{center}
 \includegraphics[angle=-90,width=0.9\linewidth]{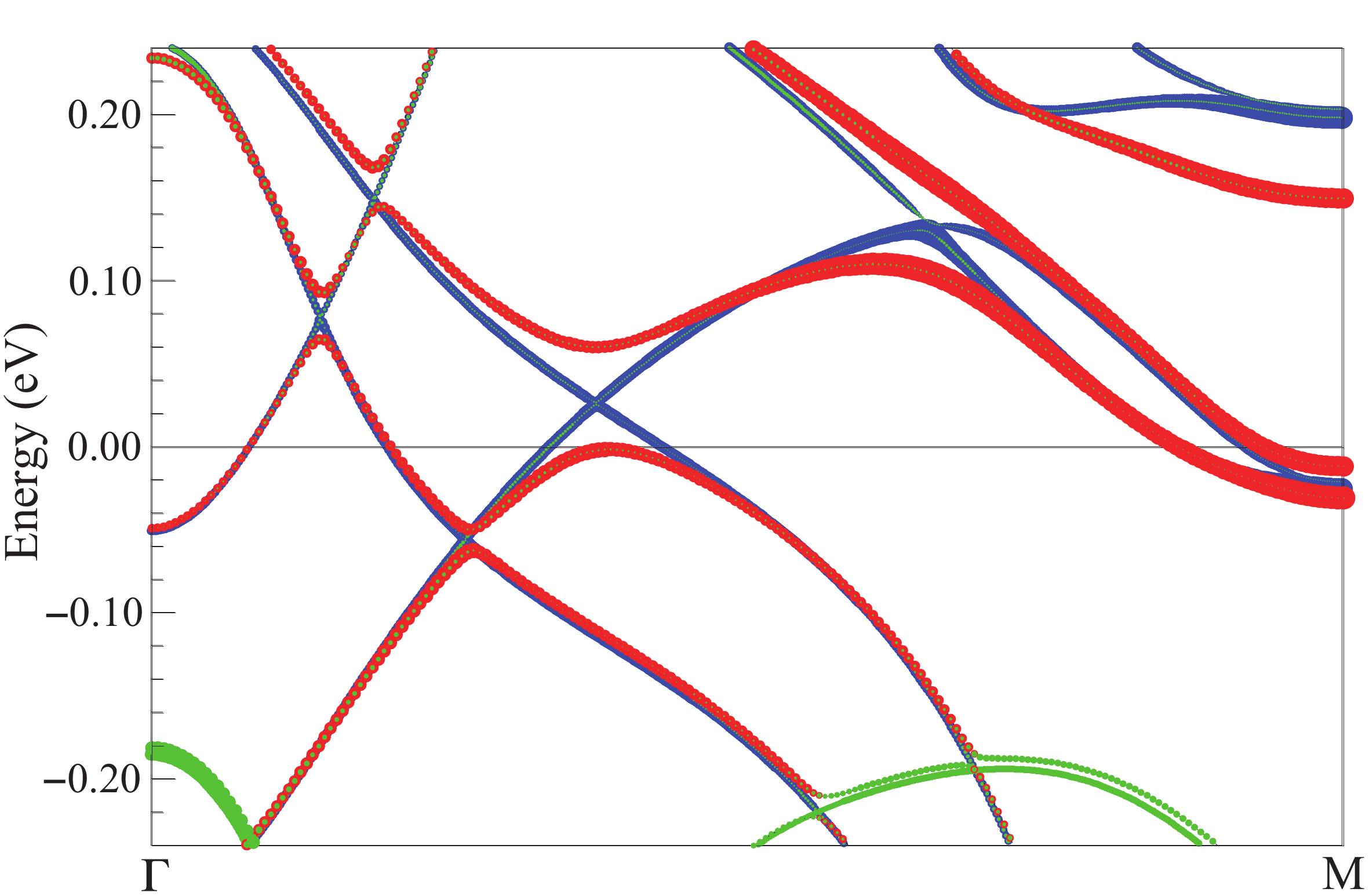} 
  \caption{
  (Color online) Enlarged view of the PM and the LMAF bands along the $\Gamma -$M ($\Sigma$) high-symmetry direction in the simple tetragonal BZ. The character of the LSDA bands is given through the color of the bands (green: Ru $4d$, red: U $5f$ in LMAF phase, blue: U $5f$ in PM phase). The amount of Ru $4d$ or U $5f$ character in the respective bands is given by the thickness of the bands. }
\label{fig:Gap-zoom}
 \end{center}
\end{figure}

An enlarged view of the PM and LMAF energy bands along the $\Gamma -$M direction is shown in Fig.\ \ref{fig:Gap-zoom}. In addition we have highlighted the orbital character of the bands through the colors and the amount of orbital character through the thickness of the bands. Ru $4d$ character is shown by the green color; the bands that consist primarily of Ru $4d$ character appear about 0.20 eV below $E_F$. The bands closer to the Fermi energy contain dominantly U $5f$ character, as is shown by the blue color in the PM phase and the red color in the LMAF phase. A small admixture of Ru $d$ character is nevertheless present. The lifting of the degenerate band crossing is clearly borne out uranium $5f$ dominated states. The bands with mainly Ru $4d$ character are unaffected. The gap opening due to the re-hybridization of states in the LMAF phase is about 60 meV wide and located in a narrow reciprocal space region at a distance of about $0.25 a^{\star} - 0.3 a^{\star}$ from the $\Gamma$ point. These values agree well with those observed in recent 
%scanning tunneling spectroscopy 
STS measurements.\cite{schmidt10}

{\urusi} is known to be a compensated metal (see, e.g., Refs.\ \onlinecite{kasahara07} and \onlinecite{ohkuni99}). The LSDA-computed energy bands of {\urusi} (Fig.\ \ref{fig:bands-123})
are fully consistent with the property, as has been pointed out recently.\cite{elgazzar09,biasini09}
Both the opening of the gap in the symmetry-broken phase and the compensated metal character are closely connected to the uranium $5f$ occupancy. Our LSDA calculations predict a $5f$ occupancy of 2.7,\cite{disclaimer} a value which is consistent with recent electron energy loss-spectroscopy (EELS) measurements.\cite{jeffries10}

\begin{figure}[tbh]
  \includegraphics[width=0.45\textwidth]{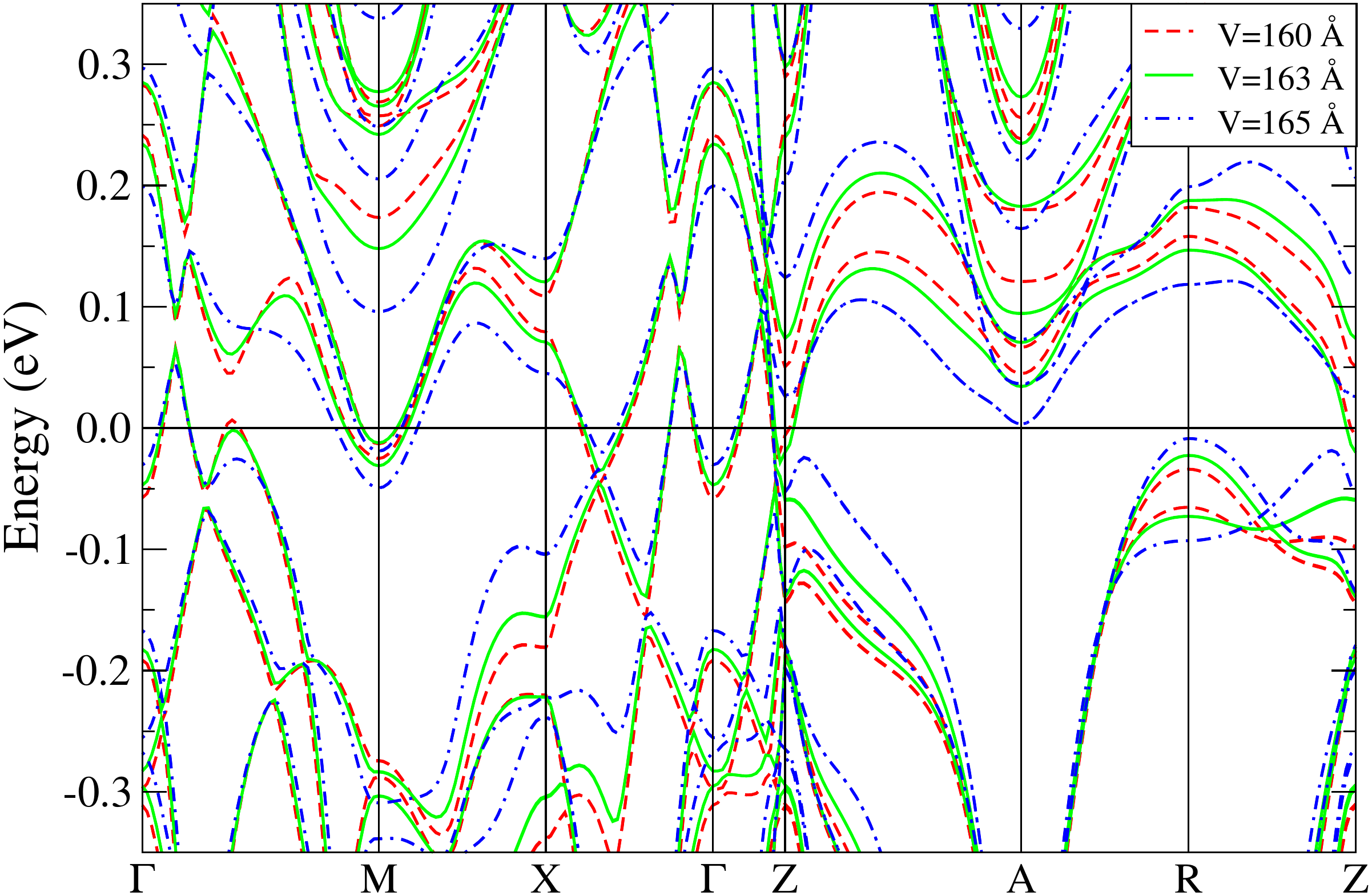}
  \caption{(Color online)  Computed LSDA energy dispersions of antiferromagnetic {\urusi} for several unit cell volumes around the experimental volume (162.9 {\AA}$^3$). 
   \label{fig:bands-V}}
\end{figure}

The lifting of a FS instability in the LMAF phase is a significant feature of {\urusi} obtained from {\it ab initio} calculations. The gap appearing around the Fermi level is narrow and might therefore sensitively depend on the lattice constants. In Fig.\ \ref{fig:bands-V} we show the influence of the volume on this feature. For a range of volumes about the experimental volume the gapping  property is found to be stable. We have similarly investigated the influence of the $z_{\rm Si}$ coordinate on the FS gapping (not shown). Also for the $z_{\rm Si}$ coordinate we find that the gapping property is stable for a range of values around the experimental one.

To end this LSDA/GGA band structure section we briefly mention that 
several LSDA electronic structure calculations have been reported for {\urusi}.\cite{norman88,rozing91,ohkuni99,yamagami00} Rozing {\it et al.}\cite{rozing91} and 
Ohkuni {\it et al.}\cite{ohkuni99} reported LDA calculations for PM {\urusi}, Yamagami and Hamada\cite{yamagami00} reported LSDA calculations of antiferromagnetic {\urusi}. The non full-potential calculations\cite{rozing91,ohkuni99} for the PM phase are in reasonable agreement with our full-potential results. Our energy bands and FS of AF {\urusi} are however distinctly different from earlier published results.\cite{yamagami00}  A reason for this difference is not known.  As mentioned before, we have verified that independent state-of-the-art electronic structure codes give nearly identical results. In Ref.\ \onlinecite{yamagami00} an AF state with nearly compensating antiparallel spin and orbital moments (possibly obtained with an orbital polarization term) is proposed as a solution for the small moment antiferromagnetic (SMAF) phase; but the SMAF phase is nowadays considered to be parasitic rather than intrinsic.\cite{takagi07}

\begin{figure}[thb]
  \includegraphics[width=0.45\textwidth]{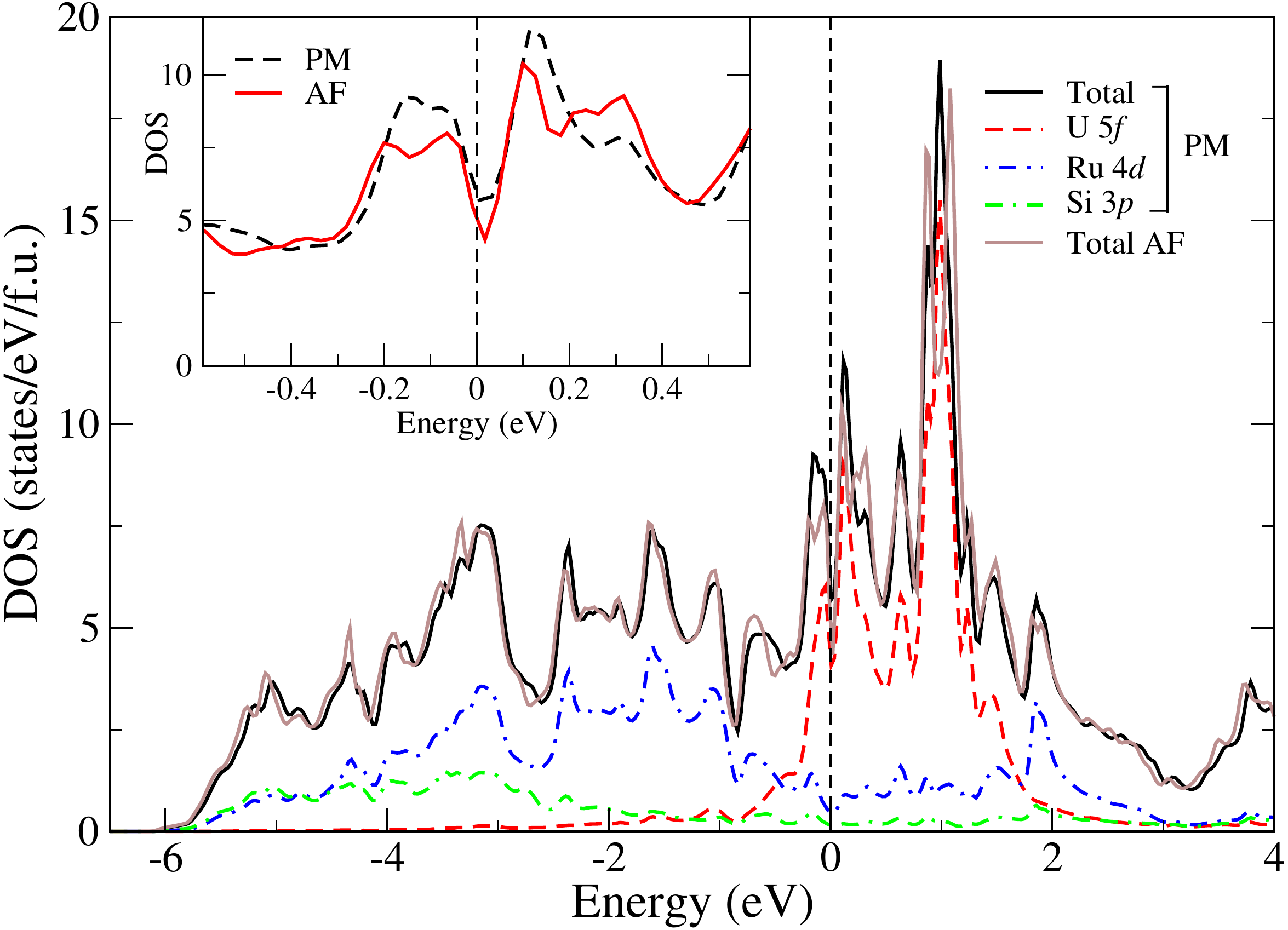}
  \caption{(Color online)  Partial density of states (DOS) of {\urusi} in the paramagnetic and large moment antiferromagnetic phases. The inset shows an enlargement of the PM and AF DOS around the Fermi energy (at 0 eV), illustrating the partial gapping occurring at the Fermi energy.
   \label{fig:DOS}}
\end{figure}

\subsubsection{Density of states}

The computed total and partial densities of states (DOSs) of {\urusi} are shown in Fig.\ \ref{fig:DOS}. Note that the Fermi level (at 0 eV)  falls precisely in a sharp minimum of the total DOS. The contribution of the uranium $5f$ states increases, starting from about 2 eV below $E_F$;  at $E_F$ and up to 1.5 eV above $E_F$, the uranium $5f$ dominate the DOS. The hybridized Ru $4d$ states extend from -6 eV to 3 eV; the hybridized Si $p$ states extend over the same energy interval. 
Yang {\it et al.}\cite{yang96}  performed a photoemission study of {\urusi}, in which they employed both He~I and He~II radiation to locate the energy position of the $5f$'s relative to the Fermi level. From the difference of the He~II and He~I emission spectra Yang {\it et al.} inferred that the U $5f$ states are  relatively delocalized and energetically extend from 1.5 eV binding energy up to the Fermi energy, where they assume a maximal contribution. The Ru $4d$ states were found to be located at about 2 eV binding energy. The computed LSDA DOS is in good agreement with these findings.
Fig.\ \ref{fig:DOS} furthermore illustrates  that the total DOSs of the PM and LMAF phases are very similar, as expected. The difference between the PM and LMAF DOSs is largest in the $5f$ energy interval; the inset of Fig.\ \ref{fig:DOS} shows the total DOS of the two phases on an enlarged energy scale close to $E_F$. Due to the gap opening on a part of the FS in the LMAF phase, the DOS minimum at $E_F$ deepens. 

%{\blue where -- The FS reconstruction ... more significant}
\subsubsection{Magnetic moments}

The total magnetic moment on the uranium atoms in the LMAF phase is reported to be 0.40 $\mu_B$ in recent measurements.\cite{amitsuka07}
Our {\it ab initio} calculations give a total moment of 0.39 $\mu_B$, in good agreement. 
The respective spin and orbital moments are $M_S = 0.36 ~\mu_B$ and $M_L = -0.75$ $\mu_B$, i.e., the orbital moment is antiparallel to the spin moment and twice as large. Detailed measurements of the separate spin and orbital moments have not been reported. Nonetheless, it can be inferred that
the computed spin and orbital moments separately are in good agreement with experiment.
Recent neutron form factor measurements\cite{kuwahara06}
 indicated a value for  $C_2= M_L / (M_L + M_S)$ of about 1.8$\pm$0.2.  From the theoretical values of $M_L$ and $M_S$ we obtain $C_2 =1.9$, implying that the predicted values for the spin and orbital moments are indeed consistent with experiment
 (which, re-calculating from $C_2$ and total moment, gives $M_S = 0.32\pm 0.04$ $\mu_B$, $M_L =-0.72 \pm 0.08$ $\mu_B$).

\subsubsection{Transport properties}

The thermal and charge transport properties of {\urusi} are experimentally well documented.\cite{palstra85,maple86,mcelfresh87,dawson89,mentink96,schoenes87,bel04,behnia05,jeffries07,kasahara07,motoyama08,hassinger08,jeffries08}
The normal\cite{palstra85,maple86,mcelfresh87,shishido09} and Hall\cite{schoenes87,dawson89,kasahara07,oh07} resistivity as well as the thermal conductivity and Nernst effect\cite{bel04,behnia05,kasahara07} display a clear signature of the HO transition, consistent with the sudden removal of a part of the FS at $T_0$. 
This distinct jump in the transport quantities is present both at the phase transition from the PM to the HO phase and that of the PM to the LMAF phase,\cite{mcelfresh87,jeffries07,motoyama08}
yet detailed charge transport measurements revealed that the FS gapping in the LMAF phase is distinctly larger than in the HO phase.\cite{mcelfresh87,jeffries07,jeffries08,motoyama08,bourdarot10} Maple {\it et al.}\cite{maple86}  expressed the FS removal in terms of the opening of a partial FS gap $\Delta$. 
The gap opening in the HO phase was measured to be about $70-80\%$ of that of the LMAF phase ($\Delta_{_{\rm HO}} \approx 75$ K, $\Delta_{_{\rm LMAF}} \approx 100$ K).\cite{mcelfresh87,jeffries08,motoyama08,bourdarot10} 

The resistivity change in the transition from the PM to the LMAF phase is accessible from the electronic structures.  To compute the electrical conductivities in these two phases, 
we have used the Kubo linear-response formulation in constant relaxation time approximation. Apart from the Fermi velocities, the conductivity expression contains an unknown electron lifetime which enters as a constant pre-factor. The electron relaxation time dependence drops out when the resistivity change is evaluated. 
For the DFT-GGA electronic structure we compute an unexpectedly large and also anisotropic resistivity change due to the opening of the gap at the PM to LMAF phase transition. 
%that the theoretical resistivity change due to the opening of a the  FS gap is unexpectedly large and also anisotropic. 
The computed resistivity jumps are $(\rho_{_{\rm LMAF}} -\rho_{_{\rm PM}})/ \rho_{_{\rm PM}} (J|| c)= 620\%$,   and   $(\rho_{_{\rm LMAF}} -\rho_{_{\rm PM}})/ \rho_{_{\rm PM}} (J|| a)= 160\%$.  
In the experiments\cite{palstra86,mentink96,zhu09} the resistivity signal is superimposed on a large background, $\sim \rho_0 + AT^2$, due to incoherent and phonon scattering. We have subtracted this background to obtain the resistivity change due to the partial FS gapping only. In this way we obtain 
the measured resistivity changes in the PM to HO phase transition, which are about 400\% and 100\% for current along the $c$, respectively, $a$ axis.\cite{palstra86} These values are consistent with the resistivity jumps computed for the LMAF phase, but they are somewhat smaller.
%The theoretical resistivity changes are larger than the experimental counterparts. 
This might be related to the fact that the measured resistivity jump pertains to the HO phase, in which the partial FS gap is smaller,  about $ 70-80\%$ of that of the LMAF phase.\cite{mcelfresh87,jeffries07,motoyama08}
Hence, the estimated experimental resistivity changes in the PM to LMAF transition would be  higher (a plain scaling would give $500\%$ and $125\%$ for $J||c$ and $J||a$, respectively).
%both for $500-570\%$ and   $125 - 143\%$, for $J||c$ and $J||a$ and 
%These values are consistent with the computed resistivity changes, but somewhat smaller, because 
We also mention that the conductivity calculations pertain to the $T=0$ K coherent electronic structure, whereas the measurements were performed in the temperature range around $T_0$.
The observed anisotropy ratio of the resistivity jump is  4\,:\,1 (Ref.\ \onlinecite{palstra86}), a value which is in very good agreement with the theoretical anisotropy ratio of 3.9\,:\,1 predicted on the basis of the DFT-GGA electronic structures.
 
Magneto-transport studies revealed that {\urusi} is a low-carrier, electron-hole compensated metal.\cite{ohkuni99,kasahara07,behnia05}
As was pointed out\cite{elgazzar09,biasini09} recently,  itinerant $5f$ calculations (LDA or GGA) indeed accurately predict this feature for {\urusi}. The electron and hole Fermi volumes in the PM phase cancel each other within $2\%$.\cite{biasini09} The number of holes has been determined from Hall effects measurements  to be 0.017$\le n_h$$ \le 0.021$ per U-atom  in the HO phase, and  0.1 per U, respectively, in the PM phase.\cite{kasahara07,oh07} 
We have used the computed intraband plasma frequency,  $\omega_p^2= \frac{4\pi e^2}{mV_{UC}} n$,  to determine the number of carriers, $n$, in both the PM and LMAF phases ($V_{UC}$ is the unit cell volume).  In contrast to the Fermi volume, the plasma frequency is a FS integral and therefore it counts only the carriers that contribute to the transport (at $T=0$~K).  The computed number of holes is 0.08/U-atom and 0.0185/U-atom in the PM phase, respectively, LMAF phase, in reasonably agreement with the experimental data. The calculated values emphasize that the FS gapping in the PM to LMAF transition strongly reduces the number of carriers by a factor of four. Hall effect measurements give that there are about five times less  carriers in the HO phase than the PM phase.\cite{oh07}

The computed $5f$ itinerant (LSDA or GGA) FS gap has recently been compared to experimental values.\cite{elgazzar09} The FS gap at the transition to the HO phase was first determined by Maple {\it et al.}\cite{maple86} from specific heat measurements.  This FS gap, averaged over the whole BZ, was estimated to be $\Delta_{_{\rm HO}} \approx 11$ meV. Somewhat smaller gaps for the HO phase of about 7 meV were obtained from transport measurements,\cite{mcelfresh87,mentink96,jeffries07,motoyama08,bourdarot10} and a larger gap of about 10 meV was measured for the LMAF phase.\cite{mcelfresh87,jeffries07,motoyama08}  
The FS gap which is predicted by DFT delocalized $5f$ calculations\cite{elgazzar09} is strongly $\boldsymbol{k}$-dependent (see Fig.\ \ref{fig:Gap-zoom} and Fig.\ \ref{fig:nesting-AFM} below). 
The LMAF gap $\Delta_{_{\rm LMAF}}$ varies from maximally 65 meV along the $\Gamma -$M ($\Sigma$)-direction to 0 meV along the $\Gamma -$X ($\Delta$) -direction. 
The larger theoretical gap obtained in certain places in the BZ is not inconsistent with the smaller BZ-averaged gaps obtained from transport measurements. Moreover, the computed gap pertains to the coherent ($T= 0$ K) electronic structure, but in the experimental analysis of the transport data  both a $\boldsymbol{k}$ and temperature independent gap is assumed constant from $T= 0$~K to $T_0$.\cite{maple86,palstra86,dawson89,mentink96,hassinger08,motoyama08} 

% compensated material;
% hall effect measurements-number of holes, jump in resistivity and anisotropy of resistivity
% MORE no of holes etc.?
% Theory needs to explain this.
% Nernst effect!!

\subsubsection{Optical spectra}

\begin{figure}[thb]
 \includegraphics[width=0.45\textwidth]{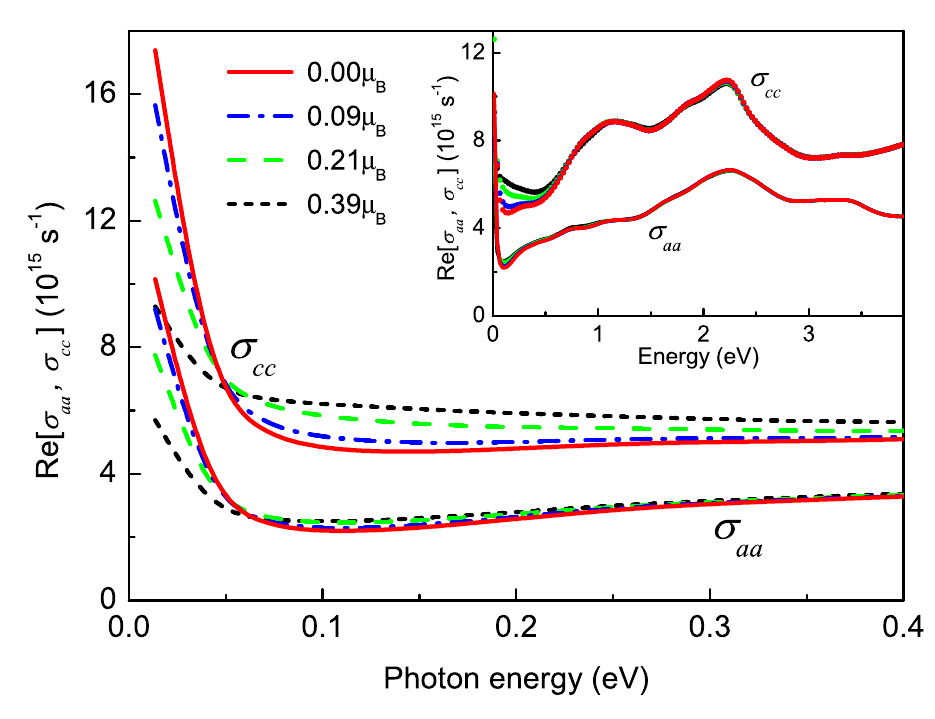}
  \caption{(Color online)  Calculated optical conductivity spectra, Re\,$[\sigma_{aa} (\omega)]$ and Re\,$[\sigma_{cc} (\omega)]$, of URu$_2$Si$_2$. Computed spectra (at $T=0$ K) are given for different uranium total magnetic moments,  starting from that of the PM phase (0.0 $\mu_B$) up to that of the LMAF phase (0.39 $\mu_B$). 
  The inset shows the computed spectra on a larger energy interval.  
  Note that the Re\,[$\sigma_{cc} $] spectra have been shifted upward by  by $3\cdot 10^{15}$ s$^{-1}$ for sake of visibility.  
  \label{fig:sigma}}
\end{figure}

The optical spectra of {\urusi} have been measured in several investigations.\cite{bonn88,thieme95,degiorgi97}
The gapping occurring in the HO phase was observed originally by Bonn {\it et al.},\cite{bonn88} more detailed infrared optical measurements of the gapping with Re doping were performed by Thieme {\it et al.} (Ref.\ \onlinecite{thieme95}) and Degiorgi {\it et al.} (Ref.\ \onlinecite{degiorgi97}); the reported spectra\cite{bonn88,thieme95,degiorgi97} are in good agreement with one another. A very recent optical study\cite{levallois10} obtained, however, a difference in the low frequency response.

Using linear-response theory we have computed the optical conductivity of {\urusi} for the two possible geometries, $E || c$ and $E || a$, where $E$ is the electric field vector of the light.
In Fig.\ \ref{fig:sigma} we show the calculated conductivity spectra, Re\,$[\sigma_{aa} (\omega)]$ and Re\,$[\sigma_{cc} (\omega)]$. The plotted optical conductivity spectra include both interband and intraband contributions and have been calculated for a range of static AF moments, going from the PM phase ($0.0~\mu_B$) up to the full LMAF phase (0.39 $\mu_B$). The theoretical spectra illustrate the effect of the increased opening of the FS gap in the vicinity of the Fermi energy.   The optical conductivities, being maximal for the PM phase near zero energy, become progressively reduced for the various static anitiferromagnetic  phases, in particular for small photon energies well below 50 meV.
The largest drop of Re\,$[\sigma (\omega)]$ is obtained for the LMAF phase, where the FS gapping is the largest. The computed behavior agrees reasonably well with experimental observations. 
Bonn {\it et al}.\cite{bonn88} measured the optical response of {\urusi} in the basal plane ($E || a$). They observed  a reduction of the reflectivity in the HO phase for photon energies below 30 meV. Our calculation predicts a drop in Re\,$[\sigma_{cc}]$ below 40 meV. Bonn {\it et al.}  did not measure  $\sigma_{cc}$  ($E || c$), but our calculations predict that a larger reduction should occur for Re\,$[\sigma_{cc}]$  at small energies well below 50 meV. 
We also note that together with the progressive FS gapping, there is a transfer of spectral weight to higher energies. Re\,$[\sigma _{aa} ]$ increases slightly above 50 meV. A larger spectral weight transfer occurs for  Re\,$[\sigma_{cc}]$ for energies of 50 meV up to 600 meV (see inset in Fig.\ \ref{fig:sigma}). Above 200 meV, respectively, 600 meV, the influence of the gapping on the optical conductivity spectra for $E || a$, respectively, $E || c$, has vanished.

The inset in Fig.\ \ref{fig:sigma} shows the computed spectra for $E || a$ and $E || c$ on a wider energy scale. An interband peak is presented just above 2 eV in both  Re\,$[\sigma_{aa}]$ and Re\,$[\sigma_{cc}]$. Experiment also detected a peak at this energy.\cite{degiorgi97} 

The experimental spectra\cite{bonn88,thieme95} reveal a particular feature which is not present in the calculated spectra. The reduction of the Drude weight at low frequencies leads to an increased spectral weight at $7 -8$ meV.\cite{bonn88,thieme95} The origin of this transferred spectral weight is currently unknown; it was not observed in a recent study.\cite{levallois10} It might nonetheless signal a difference between the experimental and computed theoretical spectra.

\subsubsection{Specific heat and magnetic entropy}

 The linear-temperature specific heat coefficient of {\urusi} in the HO phase is, with about 50 mJ/mol\,K$^2$,\cite{palstra85,maple86,schlabitz86,fisher90} not particular high, implying that {\urusi} in this phase is not a heavy-fermion material. The Sommerfeld coefficient is comparable to that of, e.g., UGa$_3$,\cite{cornelius99} which is an itinerant antiferromagnet.\cite{nakamura02,rusz04}
 The unrenormalized specific heat coefficient calculated with the LSDA approach is about 9 mJ/mol\,K$^2$, i.e., there is an expected mass renormalization of six, a value not unusual for actinides. As a consequence, the computed LSDA bands will become renormalized,  but not strongly. Our LDA+DMFT calculations (to be presented below) indicate a further influence of the dynamic part of the electronic self-energy $\Sigma (\omega )$, through which a renormalization of the bare LSDA band masses would occur. However, as we can currently not compute Re\,[$d \,\Sigma (\omega ) / d \omega $] down to low enough temperatures, we refrain from giving values for the estimated mass renormalization. Also, low-energy spin-fluctuations, which are not accounted for in the bare specific heat coefficient, can be expected to give a considerable enhancement.\cite{vandijk97}
 
The entropy of {\urusi} has drawn attention from the beginning.\cite{maple86,schlabitz86,fisher90} The phase transition to the HO state was originally discovered from a $\lambda$-type anomaly in the specific heat;\cite{palstra85,maple86,schlabitz86} the related magnetic entropy change in the $\lambda$ anomaly is, with about 0.16\,$R$\,ln\,2, relatively large.\cite{maple86,schlabitz86,fisher90} Such entropy removal can, in particular, not be explained\cite{buyers96} by assuming a phase transition to a small moment antiferromagnetic (SMAF) state that at first was thought to be connected to the HO transition.
\cite{broholm87,isaacs90,mason90,broholm91} 

The total magnetic entropy $S_{m}$ has been determined by van Dijk {\it et al.}\cite{vandijk97} and Janik,\cite{janik08} through subtracting the measured specific heat of ThRu$_2$Si$_2$, which has no $5f$ electrons, from that of {\urusi}. 
From the specific heat difference  a total electronic entropy $S_{m} (T) =
\int_0^{T} (\Delta C/T') dT' $ approaching $R$\,ln\,4 mJ/mol\,K was obtained.\cite{janik08} This value is not inconsistent with our LSDA calculations, predicting a low-temperature $5f$ count of 2.7. Assuming at higher temperatures an occupancy of three $5f$ electrons, a spin entropy of $R$\,ln\,4 (i.e., $R$\,ln\,$(2S+1)$, with $S=3 \times 1/2$) follows. 

For the HO phase, the total electronic entropy at $T_0$
%, $S_{m} =\int_0^{T_0} (\Delta C/T) dT$,  
amounts to about 0.25\,$R$\,ln\,2 (Ref.\ \onlinecite{fisher90}).
As was pointed out several times, assuming the opening of a gap $\Delta$ in the electronic spectrum, the electronic specific heat would scale as $C_{m} (T) \propto {\rm exp}( - \Delta/ k_B T)$, which fits the measured specific heat in the HO phase extremely well. \cite{maple86,fisher90,vandijk97}
In fact, the opening of gap in the magnetic excitation spectrum can {\textit{wholly}} explain the entropy removed at the HO transition.\cite{vandijk97,wiebe07}
The magnitude of the gap was estimated to be about 11 meV from specific heat measurements,\cite{maple86,fisher90} or smaller from inelastic neutron measurements.\cite{wiebe07,santini00,bourdarot10}
Our energy band calculations also show that the HO gapping removes a considerable amount of accessible states at $E_F$. The bandstructure gap computed here is in fact larger, maximally 60 meV (for the LMAF phase), but it is strongly $k$-dependent. The $k$-averaged gap would thus be considerably smaller and be consistent with the entropy loss associated with the HO transition.

\subsection{LSDA+$U$ and $5f$-core calculations}

\begin{figure}[tbh]
  \includegraphics[width=0.45\textwidth]{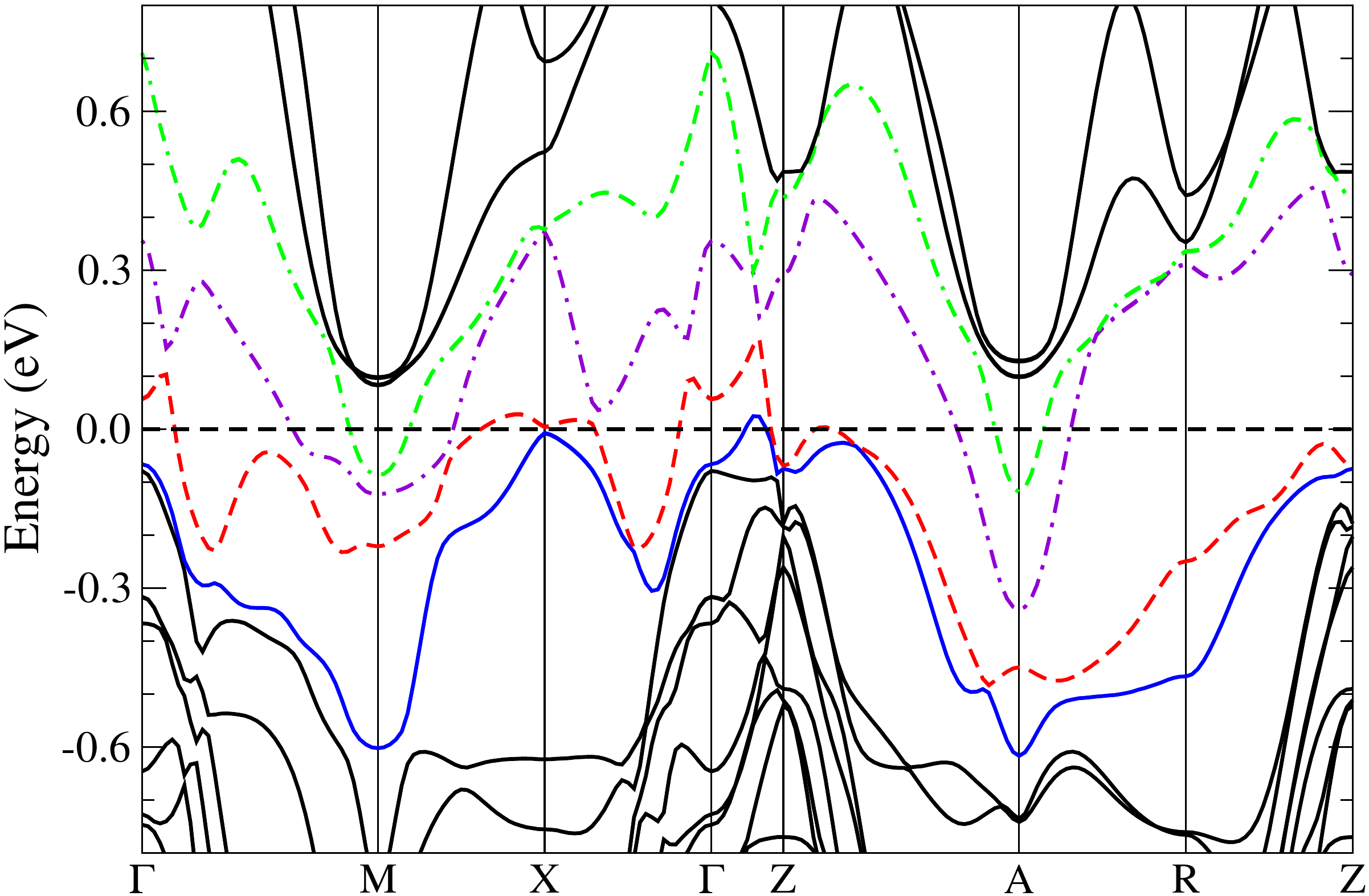}
  \caption{(Color online)  Energy dispersions of {\urusi} in the LMAF phase, computed with the around-mean field GGA+$U$ approach, with $U$=1.4 eV and $J$=0.68 eV. The bands crossing the Fermi level are high-lighted.
   \label{fig:bands+U}}
\end{figure}

Actinide materials with enhanced  Coulomb correlations between the $5f$ electrons can be computationally treated with LSDA+$U$ or GGA+$U$ calculations (see, e.g., Refs.\ \onlinecite{laskowski04}, \onlinecite{shick05}, and \onlinecite{ghosh05}), which are expected to give a good description for materials with a moderate degree of $5f$ localization.  Actinide or lanthanide materials with localized $f$ electrons are conversely well described by open $f$-core calculations, in which the $f$'s are treated as unhybridized core electrons (see, e.g., Ref.\ \onlinecite{elgazzar10}).

The energy bands of AF {\urusi} computed with the ``around-mean field" GGA+$U$ approach are shown in Fig.\ \ref{fig:bands+U}. For the Coulomb $U$ and exchange $J$ parameters we have chosen the values $U=1.4$ eV and $J=0.68$ eV. The $U$ value can be considered as relatively small and has been chosen such in order not to depart much from the LSDA solution. Nonetheless, the computed bands in Fig.\ \ref{fig:bands+U} reveal that the bands near the Fermi level are modified to a considerable extent so that also  the FS becomes quite different. The bands near the M point are pushed down, whereby new FS sheets appear. Bands near the X and $\Gamma$ points are pushed upwards, whereby also a new FS sheet appears around X. Furthermore, two new electron pockets appear around A. The gap features along the $\Gamma - $M and $\Gamma -$X directions are strongly affected; the gapping occurring along $\Gamma - $M has practically vanished. 
As we shall see below, experiments support in fact the Fermi surface predicted by LSDA calculations. This illustrates that the Fermi surface and its gapping is rather sensitive to the Coulomb $U$ in GGA+$U$ calculations. This is understandable, as the FS gap is quite small (several tens of meV) and the opening of the FS gap is due to a subtle hybridization change of $5f$ bands just above and below the Fermi level. The Coulomb $U$ acting on the $5f$ states changes the $5f$ band dispersions 
%already 
substantially. 

\begin{figure}[thb]
  \includegraphics[width=0.45\textwidth]{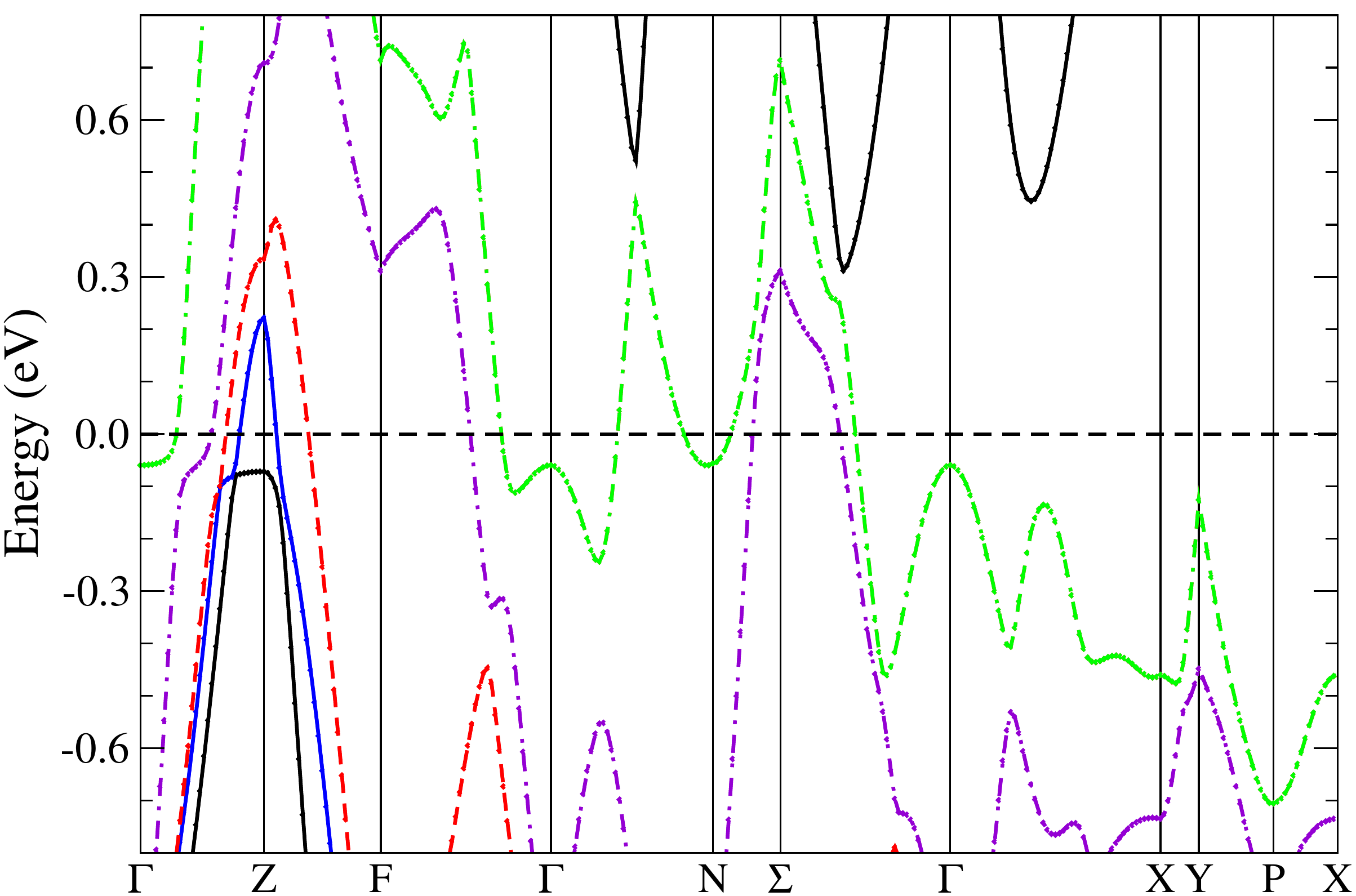}
  \caption{(Color online)  Energy dispersions of non-magnetic {\urusi} computed with a localized $5f^2$ configuration. The high-lighted bands are the ones crossing the Fermi level. The used high-symmetry points in the \textsc{bct} BZ are indicated in Fig.\ \ref{fig:BZ}.
   \label{fig:bands-core}}
\end{figure}

As mentioned before, a large number of theories for the HO of {\urusi} are based on the assumption of completely or nearly localized $5f$ electrons.\cite{santini94,kiss05,fazekas05,ohkawa99,okuno98, hanzawa05,hanzawa07,haule09,harima10,haule10}
Particularly, an underlying localized $5f^2$ configuration has been discussed recently.\cite{kiss05,hanzawa05,hanzawa07,haule09,harima10,haule10}
In itself, a localized $5f^2$ configuration possesses very interesting properties, as is can sustain both a non-magnetic spin-singlet and a magnetic triplet configuration, something which might be related to the occurrence of two different phases. 

In Fig.\ \ref{fig:bands-core} we show the energy dispersion computed for paramagnetic {\urusi} with the open-core approach for a localized $5f^2$ configuration. As expected, the $f$-core energy bands are very different from the bands obtained for PM {\urusi} assuming itinerant $5f$ valence states. Such band structure of {\urusi}, computed with WIEN2k in the PM phase was reported already in Ref.\ \onlinecite{biasini09} and is therefore not repeated here. The $f$ core energy dispersions are indeed so different from the $5f$ delocalized ones, that it makes no sense to compare them. 
As mentioned before, the $5f$ occupancy obtained from LSDA itinerant $5f$ calculations is about 2.7. Even with a small dependence on the used muffin-tin sphere radius, this occupation number is not 
%close to
 two. Hence, it is understandable that very distinct energy dispersions emerge.
The concomitant FS's are consequently also very different, as will be exemplified below when we discuss the FS of {\urusi} in detail.

LSDA-$5f^2$ core calculations were recently also performed by Haule and Kotliar,\cite{haule09}
who only show a small reciprocal space section of the bands in a narrow energy interval near the Fermi level and around the $\Gamma$-point, yet their results agree with our full-potential results. In particular, there is one band at the $\Gamma$ point near $E_F$ that has an inverted parabolic shape, and there are two bands with a steep dispersion crossing the Fermi level between $\Gamma$ and $\Sigma$. 

Whether or not a localized $5f^2$ picture is more appropriate for {\urusi} has to be considered in the light of all available experimental data. The delocalized $5f$ picture provides a quite accurate description of the known experimental data; this cannot be said of the  localized $5f^2$ configuration. In first instance one might think that the LMAF phase might be related to a magnetic, localized $5f$ configuration, but the delocalized $5f$ approach is thus far the only one that has provided an accurate explanation of the LMAF phase, which is a conventional antiferromagnetically ordered state without mysterious properties. 
%Also, the linear-temperature specific heat coefficient of {\urusi} in the HO or LMAF phase is, with about 50 mJ/mol~K$^2$,\cite{palstra85,maple86,schlabitz86} not particular high, implying that {\urusi} in these phase is not a heavy-fermion material. The Sommerfeld coefficient is comparable to that of, e.g., UGa$_3$.\cite{cornelius99}

\subsection{DMFT calculations}

 \begin{figure}[tb]
 \begin{center}
   \includegraphics[width=0.9\linewidth]{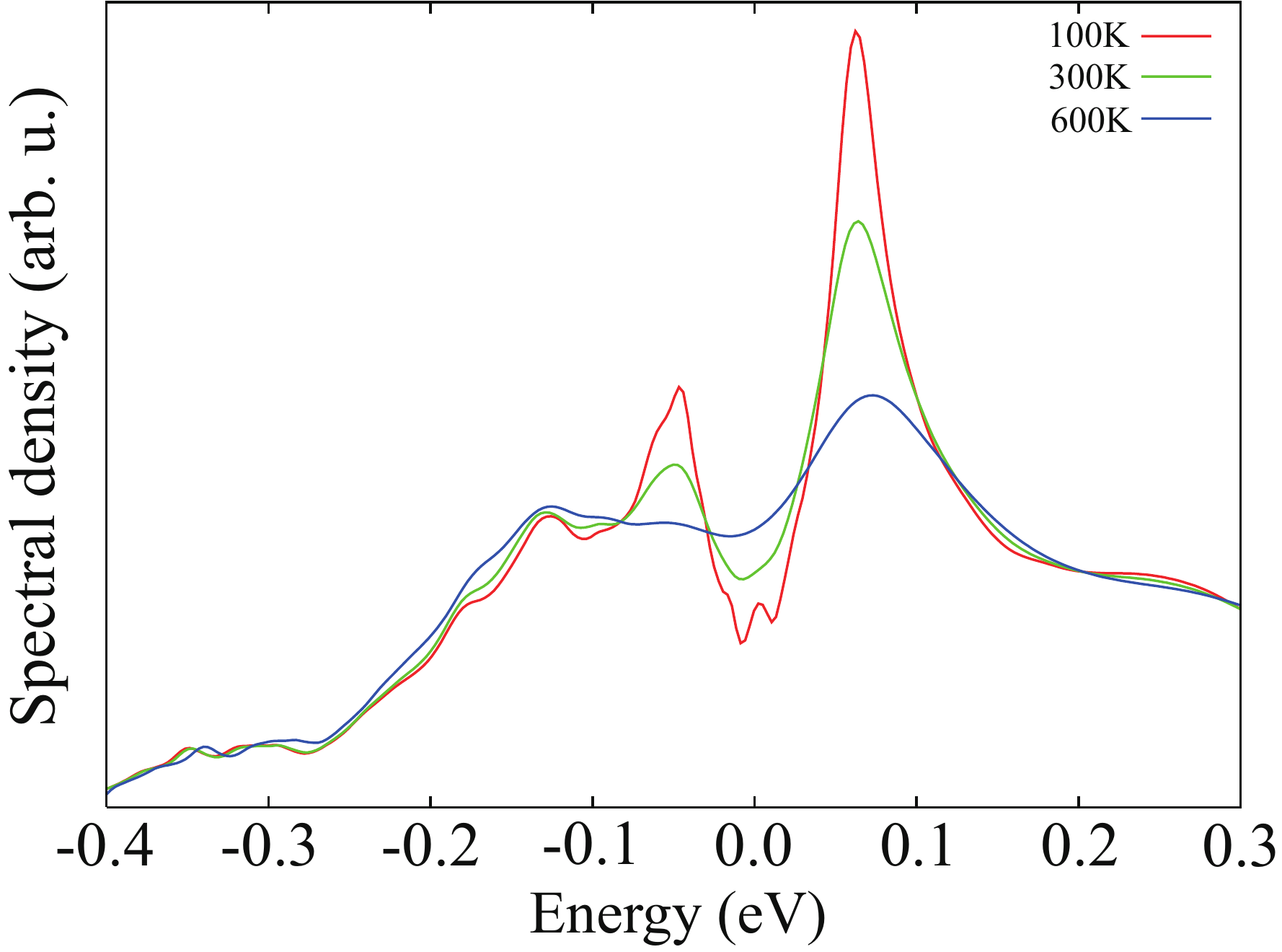}
  \caption{(Color online) The $k$-integrated quasi-particle DOS calculated with the LDA+DMFT approach for {\urusi} in the high-temperature, non-magnetic phase.}
\label{fig:DMFT-int}
 \end{center}
\end{figure}

In our LDA+DMFT calculations we used a large number of Matsubara frequency points (8192) when taking the sum over the frequencies on the temperature axis, nonetheless,  we can only compute a finite number of frequency points and therefore the calculations are valid for moderately high temperatures in practice (100 K and above). This implies that we can investigate the influence of dynamical electron configuration 
%self-energy 
fluctuations in the paramagnetic phase. At high temperatures the uranium $5f$ moments are expected to behave as incoherent, local moments. 
Note that the single-ion Kondo temperature is estimated to be  370~K in \urusi.\cite{schoenes87}
With reducing temperature, lattice coherence between the $f$ moments develops below 100~K, leading to a coherence temperature $T^{\star}$ of about 70~K, which is witnessed by a maximum in the normal and Hall resistivity.\cite{palstra85,schoenes87}
Below the coherence temperature $T^{\star}$ the $5f$ local magnetic moments are incorporated into the conduction electron sea, which greatly enhances the electron effective masses and, 
for conventional Kondo lattice materials, is expected to enlarge the Fermi surface, too.

In the LDA+DMFT calculations  we started from the LDA+$U$ approach to compute Kohn-Sham states that are subsequently 
used in the  DMFT self-consistency loop. In the LDA+$U$ part we assumed effective $U$ values  of 0.4 eV and 0.6 eV (in both cases, $J$ was set to 0.0 eV).  These  $U$ values are chosen to approximate the more localized behavior of the $5f$'s that is anticipated at higher temperatures. 
 
 In Fig.\ \ref{fig:DMFT-int} we show the computed quasi-particle density of states of {\urusi} for several temperatures. Pronounced changes in the quasi-particle DOS occur around the chemical potential (at 0 eV). Lowering of the temperature and increase of electron coherence leads to the typical opening of a quasi-particle coherence gap (also called hybridization gap) of about 100 meV. Concomitant with the opening of the quasi-particle gap, there is a build-up of spectral weight on both sides of the gap. The development of coherence gaps has been observed with infrared optical spectroscopy for several $f$-electron materials,\cite{dordevic01}
 but for {\urusi} this property has not yet been reported.

\begin{figure}[tb]
 \begin{center}
  \includegraphics[width=1.0\linewidth]{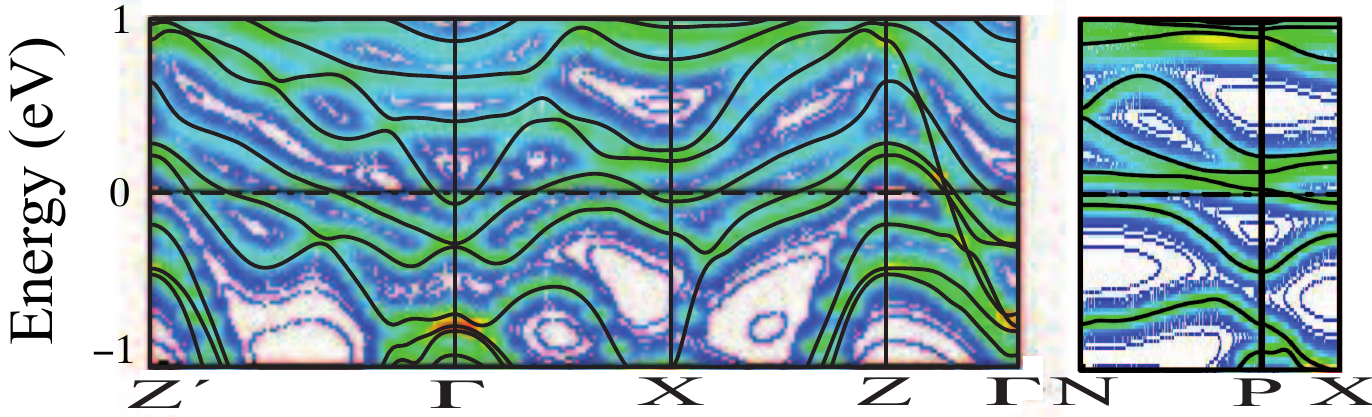}
  \caption{
  (Color online) The LDA+DMFT quasi-particle bands of {\urusi}  at $T$=100 K (for details, see text). The colors indicate the magnitude of the $k$-dependent spectral function,
  -$\frac{1}{\pi}{\rm Im}\ G ({\mbox{\boldmath$k$}},E) $. Black lines show the LDA energy bands for comparison. Note that Z$^{\prime}$ denotes the Z-point in the neighboring \textsc{bct} Brillouin zone. }
\label{fig:DMFT-bands}
 \end{center}
%\vspace{-0.8cm}
\end{figure}

 In Fig.\ \ref{fig:DMFT-bands} we show the calculated quasi-particle bands of {\urusi} at $T=$ 100 K. The bright colors depict  high intensity of the spectral function. For comparison, the non-magnetic LDA bands are shown by the  black lines. Note that the Z$^{\prime}$ point on the reciprocal space abscissa is positioned in the neighboring BZ. We observe that the LDA+DMFT quasi-particle bands are relatively close to the LDA bands.
 %(dynamical screening tends to decrease the influence of the $U$). 
 Their similarity is even more so for energy bands below -1 eV (not shown here), because these bands posses less  uranium $f$ character. Some differences between the LDA and quasi-particle bands can nonetheless be seen from Fig.\ \ref{fig:DMFT-bands}. In the N$-$P$-$X panel the quasi-particle band just above $E_F$ moves distinctly closer to $E_F$ and becomes flatter. Near the X point the quasi-particle band below the Fermi level moves slightly upwards and disperses stronger downwards towards the $\Gamma$ point. The $k$-dependent quasi-particle DOS  gives the impression of a band dispersing downwards from above the X point towards the $\Gamma$ point and crossing $E_F$ between the two points. 
  
Another DMFT calculation for {\urusi} has been reported recently by Haule and Kotliar.\cite{haule09} We note that our DMFT results are distinctly different from those of Ref.\ \onlinecite{haule09}. We have performed  self-consistent DMFT calculations using the spin-polarized FLEX impurity solver, starting from LDA+$U$ results, which should be valid for the weakly correlated uranium $f$ electrons at higher temperatures. The DMFT calculations of Ref.\ \onlinecite{haule09}, on the other hand, used the one-crossing approximation solver together with a nearly localized uranium $5f^2$ configuration. 
The difference can be understood to arise from the $5f$ configuration used in the LDA+$U$ band structure part of the LDA+DMFT calculation, which obviously determines largely the gross electronic structure, whereas the DMFT part involving dynamical self-energy fluctuations induces mainly modifications of energy dispersions in the vicinity of the chemical potential.

\begin{figure}[tb!]
%\begin{minipage}
 %\flushleft
   \includegraphics[width=0.47\textwidth]{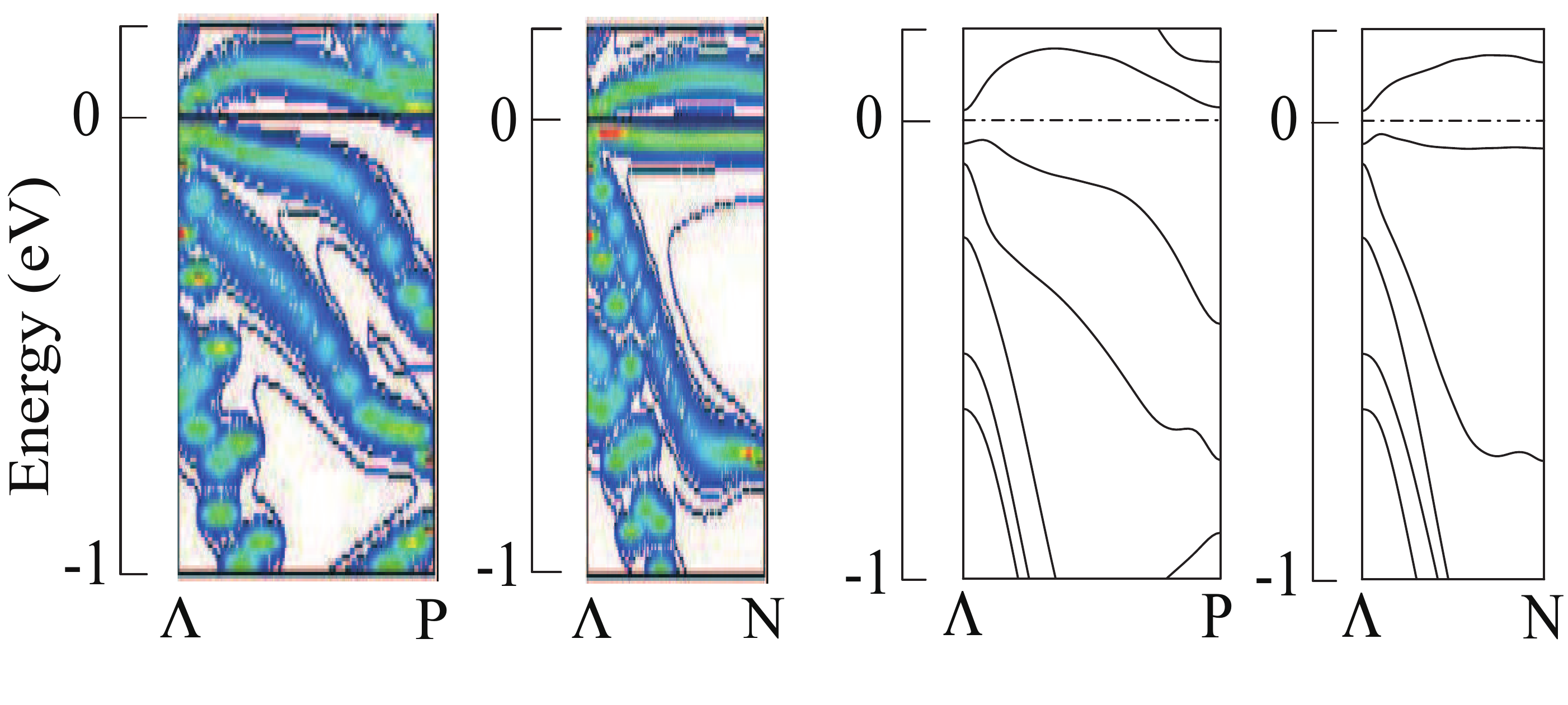}
%  \includegraphics[width=0.47\textwidth]{DMFT+bands-L2-copy.eps}
%  \end{minipage}
  \caption{(Color online) Left: LDA+DMFT quasi-particle bands for the directions $\Lambda -$N and $\Lambda -$P in the \textsc{bct} BZ ($\Lambda=$Z/2), and right:
  the LDA energy bands for the same directions.
  \label{fig:Lambda}}
\end{figure}

 DMFT $k$-dependent spectral functions can be compared to ARPES data. Several ARPES measurements on {\urusi} have been reported.\cite{ito99,denlinger01,santander09}  Ito {\it et al.}\cite{ito99} and Santander-Syro {\it et al.}\cite{santander09} both used a He~I light source, whereas Denlinger {\it et al.}\cite{denlinger01} used tunable synchrotron radiation.
Only the recent experiment of Santander-Syro {\it et al.} measured quasi-particle bands below $T_0$.
These latest measurements indicate the existence of a narrow band just below $E_F$ in the HO phase, as well as of an inverted parabolic band at $k=0$ below $E_F$; the latter band was attributed to a surface state.\cite{santander09}
In our $5f$-itinerant LSDA calculation there is no such inverted parabolic band at the $\Gamma$ point, but LDA localized $5f^2$ electron calculations do predict such a band (see Fig.\ \ref{fig:bands-core}), as do also the recent LDA+DMFT calculations of Haule and Kotliar.\cite{haule09} As photoemission at this energy is very surface sensitive, it could thus be that this band stems from a $5f$-localized response of uranium atoms on the surface.
Further investigations are therefore needed to definitely establish the origin of the inverse parabolic band.
A bulk, flat band just below $E_F$ at the $\Gamma$ point is not predicted by our delocalized $5f$ LDA or LDA+DMFT calculations.   We note, however, first, that the LDA and LDA+DMFT bands have a significant dispersion along the $k_z$ direction, and second, that the $k_z$ position in the BZ will, in an normal-emission ARPES experiment, depend on the energy of the used radiation.
 With He~I radiation, a $k_z$ position between $\Gamma$ and Z in the \textsc{bct} BZ will be probed,\cite{ito99} probably being closer to $\Gamma$ than Z. 
In Fig.\ \ref{fig:Lambda} we present computed LDA and LDA+DMFT bands for the midpoint, $\Lambda =$Z/2, between  $\Gamma$ and Z, and going to the P, respectively, N, high-symmetry points in the \textsc{bct} BZ (see Fig.\ \ref{fig:BZ}).
The plotted bands illustrate that a flat band exists in the P$-\Lambda -$N plane, just below $E_F$, being mostly flat near  $k_x$, $k_y =0$. 
At his point it is still too early to decide whether this computed flat band does or does not correspond to the observed ARPES structure.\cite{santander09} High-resolution ARPES measurements with tunable photon energy will be required to reveal the full electronic dispersions in the HO phase.

\subsection{The Fermi surface of {\urusi}}

\subsubsection{Nesting vectors}

  \begin{figure}[tb]
 \begin{center}
   \includegraphics[width=0.7\linewidth]{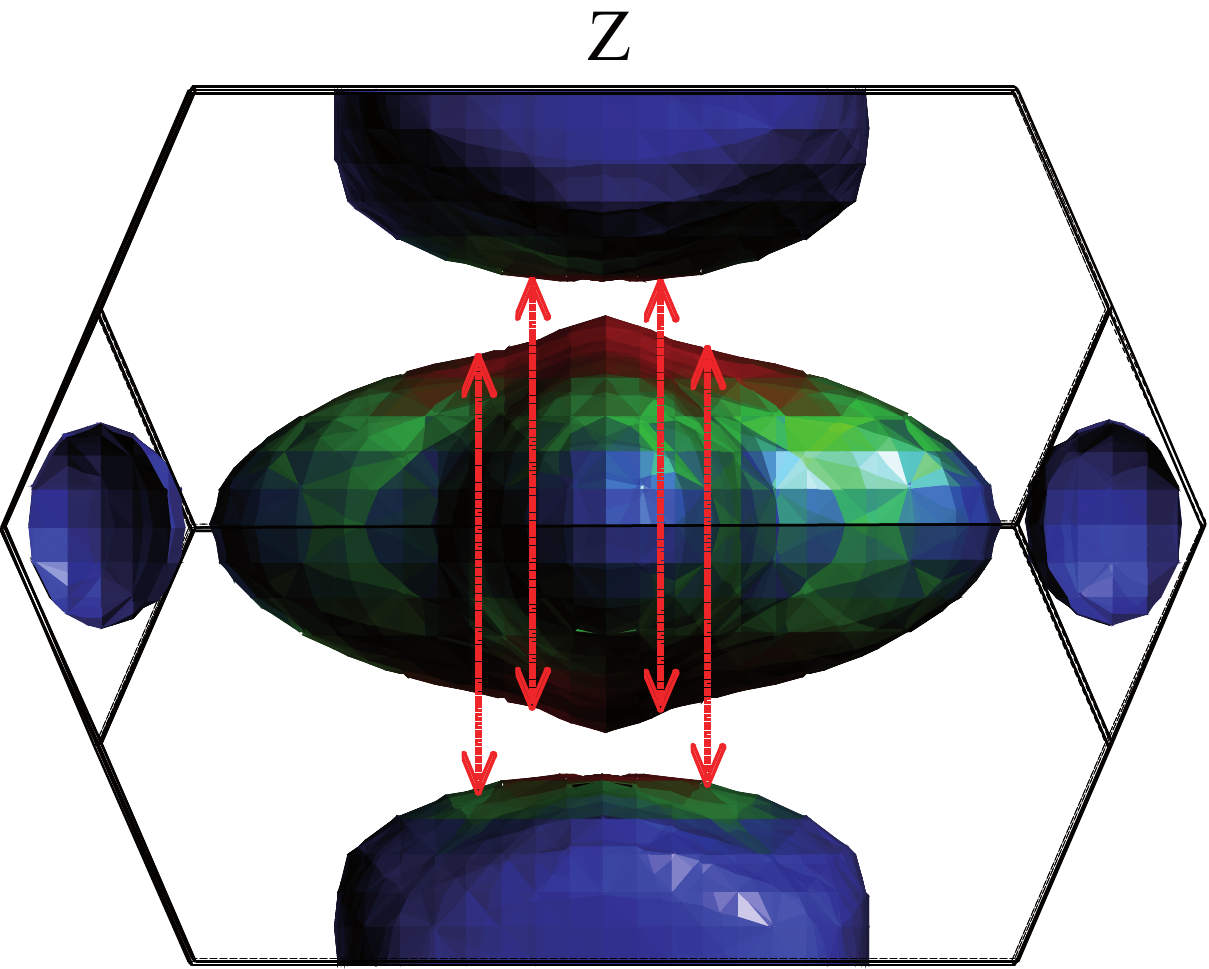}
  \caption{
  (Color online) Side-view of the Fermi surface of PM {\urusi} along the $k_x$ or $k_y$ axis of the \textsc{bct} BZ (see Fig.\ \ref{fig:BZ}). The arrows indicate the AF nesting vector (with length $2\pi/c$) connecting the two FS sheets. Note that two smaller FS sheets (one Z centered ellipsoid and one small $\Gamma$ centered ellipsoid) are not seen here.}
\label{fig:PM-nesting}
 \end{center}
%\vspace{-0.8cm}
\end{figure}

An appropriate description of the Fermi surface topology of {\urusi} is an indispensable ingredient for unraveling the nature of the HO phase as well as the unconventional superconductivity (SC).\cite{oh07,yano08,kasahara07,kasahara09}
Experimental information regarding the FS of {\urusi} has been gained from nesting vectors, identified through inelastic neutron experiments,\cite{broholm91,wiebe07,janik09,villaume08} and through extremal FS orbits, obtained from quantum-oscillation experiments.\cite{ohkuni99,jo07,shishido09}

To start our discussion, we show a
a side-view of the FS of PM {\urusi}, computed with the LSDA approach, in Fig.\ \ref{fig:PM-nesting}. The two FS sheets reveal  the existence of a nesting vector with length $c^{\star}=2\pi/c$ (i.e., half the distance from one $\Gamma$ point to the next nearest $\Gamma$ point). These two FS sheets have a similar round curvature, favorable for nesting, with the exception that close to Z/2 the $\Gamma$-centered sheet has a more pointed part, with only a  small area that would not be favorable for nesting. This FS part corresponds to a small part at the Z point in the simple tetragonal cell, which we believe to be insignificant.\cite{elgazzar09}
The identified nesting vector fits
accurately to the antiferromagnetic wavevector $\boldsymbol{Q}_{AF}= (0,~0,~1)$ of longitudinal spin-fluctuations observed in the HO phase with inelastic neutron scattering experiments,\cite{broholm87,wiebe07,villaume08} and it is the AF ordering vector of the LMAF phase.\cite{broholm91,villaume08}
This nesting vector is important for understanding the low temperature behavior of {\urusi}. When a coherent state emerges at temperatures sufficiently below the coherence temperature $T^{\star}$, the system develops a FS sustaining this nesting vector, which is favorable for AF spin-fluctuations in the HO phase and formation of long-range AF order in the LMAF phase.
Inelastic neutron experiments\cite{villaume08} showed that
the inelastic response at $\boldsymbol{Q}_{AF}$ in the HO phase becomes the static AF Bragg peak of the LMAF phase.\cite{villaume08}
% i.e., this wavevector corresponds to the long-range AF ordering. 
The thereby induced symmetry-breaking implies a folding of the \textsc{bct} BZ at Z/2, i.e., folding Z to $\Gamma$.

 \begin{figure}[tb]
 \begin{center}
 \includegraphics[width=0.8\linewidth]{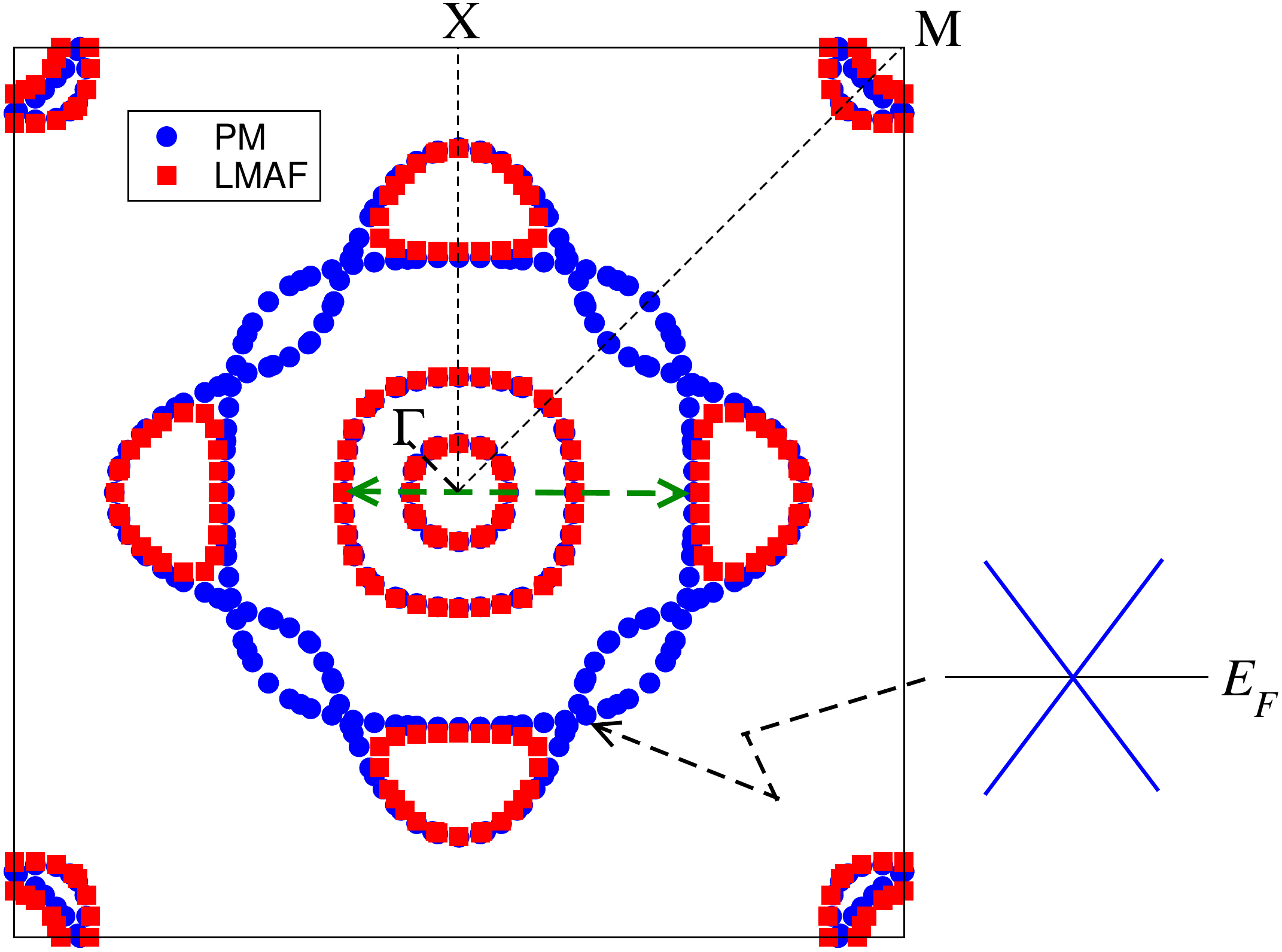}
  \caption{
  (Color online) Cross-sections of the PM (circles) and LMAF (squares) Fermi surface in the $z=0$ plane of the simple tetragonal BZ. The Fermi surface portions along the $\Gamma -$M directions are gapped in the PM to LMAF transition. Note that other Fermi surface parts are completely unaffected; the LMAF Fermi surface cross-sections are on top of the PM ones. The thick dashed arrow indicates the nesting vector $0.4\,a^{\star}$,
  % ($a^{\star}=2\pi /a$),
the thin dashed arrow indicates the position of one of the eight degenerate band crossings precisely at the Fermi energy $E_F$.  }
\label{fig:nesting-AFM}
 \end{center}
\end{figure}

A second, incommensurate nesting vector of {\urusi} has been detected at $\boldsymbol{Q}_1 = (1\pm 0.4,~0,~0)$.\cite{broholm91,wiebe07,villaume08} This nesting vector has been observed in 
{\it both} the HO and LMAF phase.\cite{villaume08}
In Fig.\ \ref{fig:nesting-AFM} we show a cross-section of the LMAF and PM Fermi surfaces in the $z=0$ plane. To draw comparison, both FS cross-sections are plotted in the simple tetragonal unit cell of the LMAF phase. As was reported recently,\cite{elgazzar09}  a clear nesting occurs (depicted by the dashed arrow) at $0.4\,a^{\star}$ ($a^{\star}=2\pi /a$), implying a nesting vector that matches precisely the measured incommensurate vector $\boldsymbol{Q}_1 = (1\pm 0.4,~0,~0)$.
 Fig.\ \ref{fig:nesting-AFM} illustrates that $\boldsymbol{Q}_1$ is a suitable nesting vector for the LMAF phase, but less so for the PM phase (shown by blue circles), because in the latter phase the FS curvature  does not support nesting as much. Hence, the incommensurate nesting vector is characteristic for the LMAF phase, but not for the PM phase. {\it A priori} we don't know how the FS in the HO phase  looks like, because this  would require knowing the order parameter of the HO, but it is known that de Haas-van Alphen experiments could not detect any notable difference between the FSs of the HO and LMAF phases.\cite{nakashima03} Likewise, inelastic neutron scattering experiments\cite{villaume08} give the same incommensurate wavevector in the HO and LMAF phases. From these facts we infer that the FS of {\urusi} in the HO phase should be quite close to the one we have computed for the LMAF phase.

The gap symmetry in the superconducting phase has been discussed recently,\cite{kasahara07,yano08,kasahara09} but the symmetry of the partial gap in the HO phase has not yet been studied.
The calculations in Fig.\ \ref{fig:nesting-AFM} illustrate that the  HO gap has in the $z=0$ plane a fourfold symmetry, in which there exist nodal lines with $d_{x^2 - y^2}$ symmetry in the \textsc{st} structure, which is equivalent to nodal $d_{xy}$ symmetry in the \textsc{bct} structure.
A possible additional phase factor in the gap structure might exist, but cannot be deduced from the current calculations.

%{\blue The good agreement between the {\it ab initio} computed and experimental nesting vectors supports the Fermi surface topology as predicted by the LSDA itinerant $5f$ approach. }

\begin{table}[h]
\begin{ruledtabular}
\begin{center}
\caption{Comparison of experimental and {\it ab initio} calculated extremal FS areas in {\urusi}, for $H\, || \,c$.
The FS cross-sectional areas are labeled according to the convention of Refs.\ \onlinecite{ohkuni99} and \onlinecite{shishido09} and have been determined with de Haas-van Alphen (dHvA) measurements \cite{ohkuni99} or Shubnikov-de Haas (SdH) measurements.\cite{jo07,shishido09} The theoretical FS orbits are shown in Fig.\ \ref{fig:FS}.
All FS cross-sectional areas are given in kT.
}
 \label{table1}
\begin{tabular}{cccccc}
  Orbit & dHvA & SdH & SdH  & & Calc. \\
          & Ref.\ \onlinecite{ohkuni99} & Ref.\ \onlinecite{shishido09} & Ref.\ \onlinecite{jo07} & &  \\
 \hline
 $\varepsilon$ &  $-$   & 1.35 &  $-$  & & 1.37 \\
 $\alpha$    &  1.05 & 0.98 & 0.85 & & 0.76 \\
 $\beta$      &   0.42 & 0.50 & 0.48 & & 0.48 \\ 
 $\gamma$, $\gamma^{\prime}$ &  0.19 & 0.25 & 0.22 & & 0.26 \\
 \end{tabular}
\end{center}
\end{ruledtabular}
\end{table}
%$^a$Ohnuki {\it et al.}\\
%$^b$Shishido \\
%$^c$Jo\\

\begin{figure}[thb]
 \includegraphics[width=0.38\textwidth]{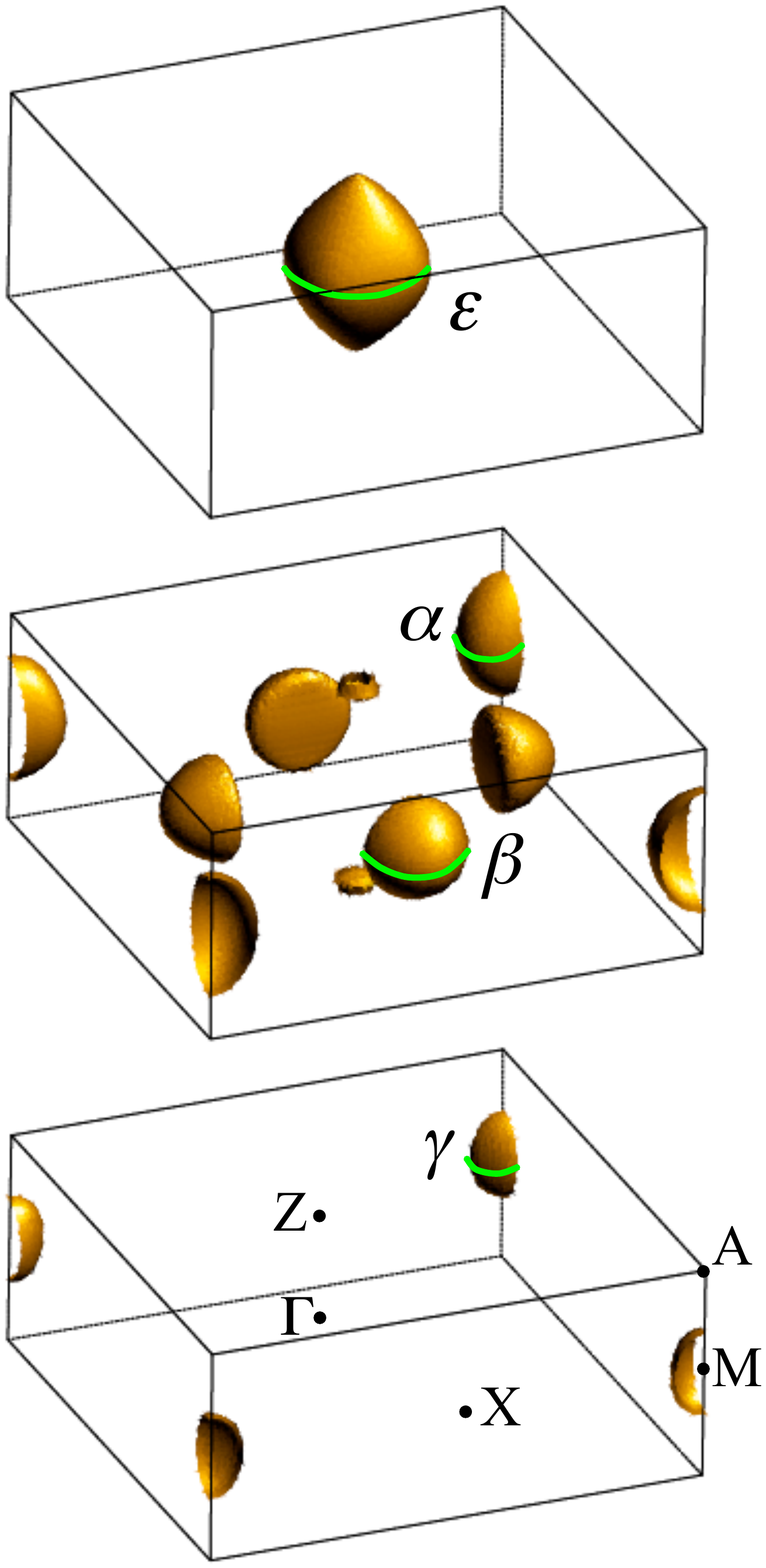}
  \caption{(Color online)  Computed Fermi surface sheets of {\urusi} in the  LMAF phase. The extremal Fermi surface orbits for field along the $z$ axis are indicated. Not shown is the small $\Gamma$-centered ellipsoid ($\gamma'$) that is inside the large $\Gamma$-centered surface in the top panel.  High-symmetry points are indicated in the bottom panel.
  \label{fig:FS}}
\end{figure}

\subsubsection{Quantum oscillations}

%Remarks regarding the Fermi surface. 
Details of the FS of {\urusi} has been investigated, too, through studies of quantum oscillations in de Haas-van Alphen (dHvA) and Shubnikov-de Haas (SdH) measurements.\cite{bergemann97,ohkuni97,keller98,ohkuni99,jo07,shishido09}
In earlier experiments,\cite{bergemann97,keller98,ohkuni97} only a few extremal orbits could be detected. The more recent dHvA and SdH measurements\cite{ohkuni99,jo07,shishido09} on purer crystals have provided a consistent set of data for  the quantum oscillations in {\urusi}. These experimental data are listed together with the calculated extremal FS cross-section areas in Table \ref{table1}. 
The dHvA experiments of Ohkuni {\it et al.}\cite{ohkuni99} as well as the SdH experiments of Jo {\it et al.}\cite{jo07} revealed three FS orbits, which these authors labeled $\alpha$, $\beta$, and $\gamma$. However, it was already noted\cite{ohkuni99} that one or more orbits were missing, because the cyclotron mass of the small FS orbits would not be able to account for the full, enhanced mass that was deduced from the specific heat. Quite recently, it was shown with SdH experiments on ultraclean crystals\cite{shishido09} that there is indeed a further branch, which was named the $\varepsilon$ branch. 
The calculated FS of {\urusi} is shown in Fig.\ \ref{fig:FS}, where we have also indicated the theoretical extremal orbits $\varepsilon$, $\alpha$, $\beta$, and $\gamma$ for magnetic field along the $z$ axis. Note that there is a small $\Gamma$-centered ellipsoid ($\gamma^{\prime}$) which is not visible in Fig.\ \ref{fig:FS} because it is masked by the large $\Gamma$-centered ellipsoid.
Table \ref{table1} illustrates that the calculated extremal orbits of the four branches are in quite good agreement with the available experiments. The computed orbits pertain to the LMAF phase, however, consistent with the observations made above, we assume the same FS for the HO and LMAF phases.  The newly discovered\cite{shishido09} $\varepsilon$  branch corresponds to the large $\Gamma$ centered rugby ball, stemming from band 107 in our calculations. The earlier reported extremal orbits correspond to the M-centered ellipsoids (branches $\alpha$ and $\gamma$), the rounded half-sphere ($\beta$), and the small, $\Gamma$-centered sphere ($\gamma^{\prime}$). The computed frequencies of these extremal orbits agree well with dHvA and SdH measurements: the extremal areas of the $\varepsilon$ and $\beta$ orbits are well reproduced. The $\alpha$ orbit is somewhat smaller than the experimental value and the $\gamma$ orbit is somewhat larger. The calculations predict, in fact, a $\gamma$ and a $\gamma^{\prime}$ orbit, which have nearly the same frequency. Also, as discussed below, the angular dependence of their cross-sectional area is very similar. This similarity appears to be coincidental. Experimentally, it would thus be difficult to distinguish between the $\gamma$ and $\gamma^{\prime}$ branches. One other difference between the experiments and the calculations is that the calculations predict an additional, small FS part around the Z-point (see Fig.\ \ref{fig:FS}.  Its extremal area is very small (about 0.07~kT), and, as mentioned previously,\cite{elgazzar09} its occurrence is related to an insignificant, small FS area at the Z-point which occurs through band folding. The origin of this small FS sheet is the pointed part of one FS sheet near $\pm$Z/2 in the PM phase (Fig.\ \ref{fig:PM-nesting}), which causes an imperfect nesting of the two FS sheets and hence an incomplete FS gapping near the \textsc{st} Z-point  when being folded to the LMAF phase.

We briefly mention that earlier studies of quantum oscillations in {\urusi} were reported by Bergemann {\it et al.},\cite{bergemann97} Ohkuni {\it et al.},\cite{ohkuni97} and Keller {\it et al.}\cite{keller98} Although the measurements were performed on less well-defined samples, there are some consistencies with the newer data.
Bergemann {\it et al.} report the observation of two orbits, which cross-sectional areas of 1.09 and 0.41 kT for magnetic field along the $c$-axis. Keller {\it et al.} reported magnetoresistive measurements from which they obtained several SdH frequencies of about 1.0 and 0.2 kT.

\begin{figure} [tb!]
\includegraphics[width=0.45\textwidth]{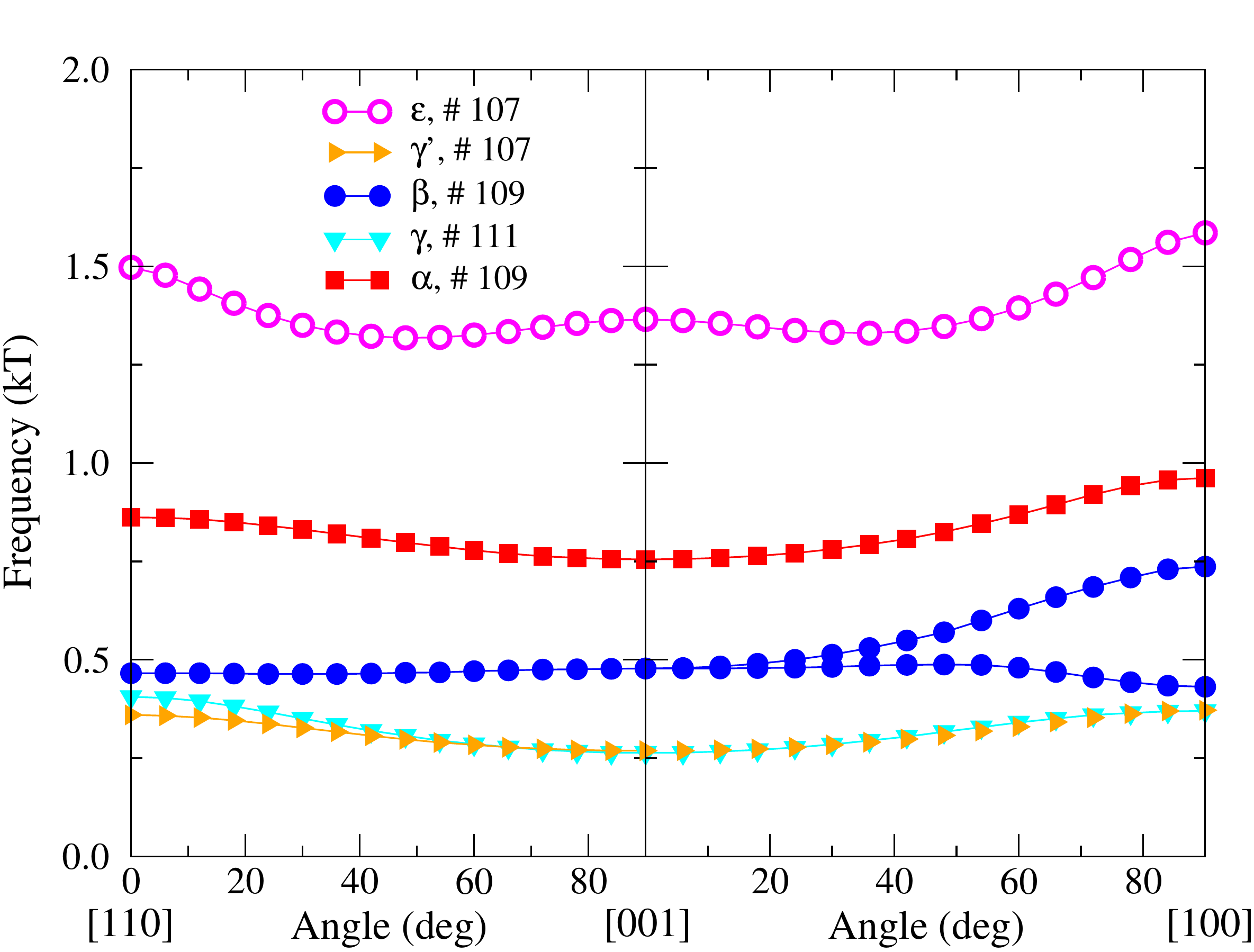}
\caption{(Color online) Calculated dependence of the extremal FS cross-sections  of {\urusi} on the magnetic field direction. The extremal orbits are visualized in Fig.\ \ref{fig:FS}. The band number from which the FS sheet originates is given after the greek label of the respective orbits.
\label{fig:dhva}}
\end{figure}

The crystallographic anisotropy of the extremal FS areas has been investigated by angular dependent dHvA and SdH experiments. The computed angular dependence of the FS areas is shown in Fig.\ \ref{fig:dhva}, for field directions in the \textsc{bct} cell. A comparison of the computed angular dependence with experimental data reveals a good overall correspondence. 
Ohkuni {\it et al.}\cite{ohkuni99} report that the $\alpha$, $\beta$, and $\gamma$ branches are quite flat, i.e., corresponding to nearly spherical FS sheets. Shishido {\it et al.}\cite{shishido09} also report that the $\alpha$ and $\beta$ branches are flat and due to spherical FS's. As Fig.\ \ref{fig:dhva} shows, the theoretical FS sheets are predicted to be relatively spherical, and hence, a very similar angular behavior is given by the calculations. Moreover, a more detailed comparison reveals that the angular anisotropy of the $\varepsilon$, $\alpha$, and $\beta$ branches is also in good agreement with experiment. The extremal frequency of the $\alpha$ orbit is predicted to increase from the [001] to [100] direction, which was also found experimentally.\cite{ohkuni99} The frequency of one part of the $\beta$ orbit decreases slightly, in accordance with experiment. 
The four half-spheres in the $z=0$ plane are anisotropic with respect to the field direction, therefore the single $\beta$ orbit for field $H||c$ splits in two orbits for fields along [100], but no splitting occurs along [110].
Such a splitting of the $\beta$ branch has been detected recently in SdH experiments.\cite{hassinger-unp}
The $\varepsilon$ orbit increases in the [100] direction, which was also observed in recent SdH measurements.\cite{shishido09} However, the SdH experiment finds a larger increase than the theory predicts. 

The crystallographic anisotropy calculated for the $\gamma$ orbit is not in accordance with experiment; the calculation predicts a slight increase of the frequency for magnetic fields towards [100], but the dHvA experiment\cite{ohkuni99} observed a slight decrease. 
Ohkuni {\it et al.}\cite{ohkuni99} mention, however, that the signal of this orbit is very weak and therefore it was difficult to follow the orbit under angular field variation. The recent SdH experiments \cite{shishido09} could not determine the angular dependence of the $\gamma$ orbit for the same reason. We also note, lastly, that the difference between the $\gamma$ and $\gamma^{\prime}$ extremal orbits is predicted to be larger for fields along the [110] direction. 

\begin{figure}[tb!]
  \includegraphics[width=0.45\textwidth]{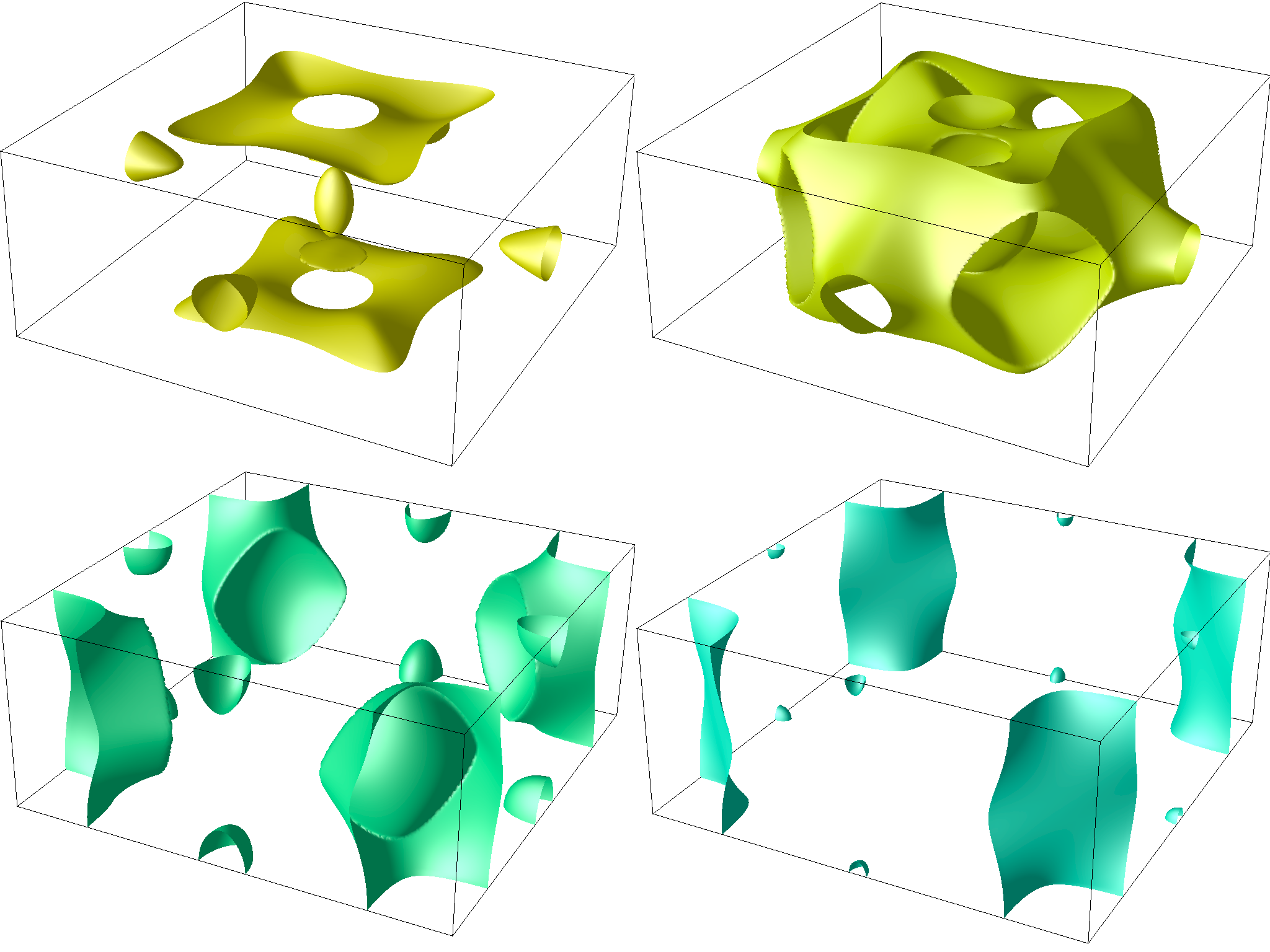}
  \caption{(Color online) The Fermi surface  of {\urusi} in the LMAF phase computed with a $5f^2$ local uranium configuration.
  \label{fig:FS-core}}
\end{figure}

As a nearly localized uranium $5f^2$ configuration has been suggested to be applicable to 
{\urusi},\cite{haule09,harima10} we briefly consider the FS that would correspond to such configuration. In Fig.\ \ref{fig:FS-core} we show the FS computed for {\urusi} in the LMAF phase obtained for a localized $5f^2$ configuration. Obviously, this FS is much larger than that computed assuming delocalized $5f$'s.\cite{note2}
 In addition, there are quite a number of extremal orbits (more than ten for $H || z$). Both these aspects of the $5f^2$ localized FS are not in correspondence with experiments. 
Conversely, the overall agreement between the experimental and theoretical FS computed with delocalized $5f$'s can be regarded as quite good, a result which definitely supports that the theoretically predicted LSDA/GGA FS is in close agreement with the experimental Fermi surface.

\subsection{Longitudinal spin-fluctuations and dynamical symmetry breaking}

\subsubsection{Potential relevance of AF spin-fluctuations}

A significant difference between the HO and LMAF phase is the existence of an intense mode of long-lived longitudinal AF fluctuations at the wavevector $\boldsymbol{Q}_{AF}$ in the HO phase, which has been identified as a fingerprint of the HO.\cite{villaume08,bourdarot10} 
This AF mode was studied in inelastic neutron experiments by Broholm {\it et al.}\cite{broholm87,broholm91}, Wiebe {\it et al.},\cite{wiebe07} and Villaume {\it et al.},\cite{villaume08} and in x-ray scattering experiments by
Bernhoeft {\it et al.}\cite{bernhoeft03,bernhoeft04} The longitudinal fluctuation was estimated\cite{broholm87,walker93} to amount to $\Delta M_z \propto 1.2~\mu_B$, its characteristic time-scale
is of the order of picoseconds or slower.\cite{bourdarot10} 
The potential importance of the longitudinal AF mode for the HO phase was recently emphasized by Elgazzar {\it et al.}\cite{elgazzar09}

\begin{figure}[tb]
  \includegraphics[width=0.45\textwidth]{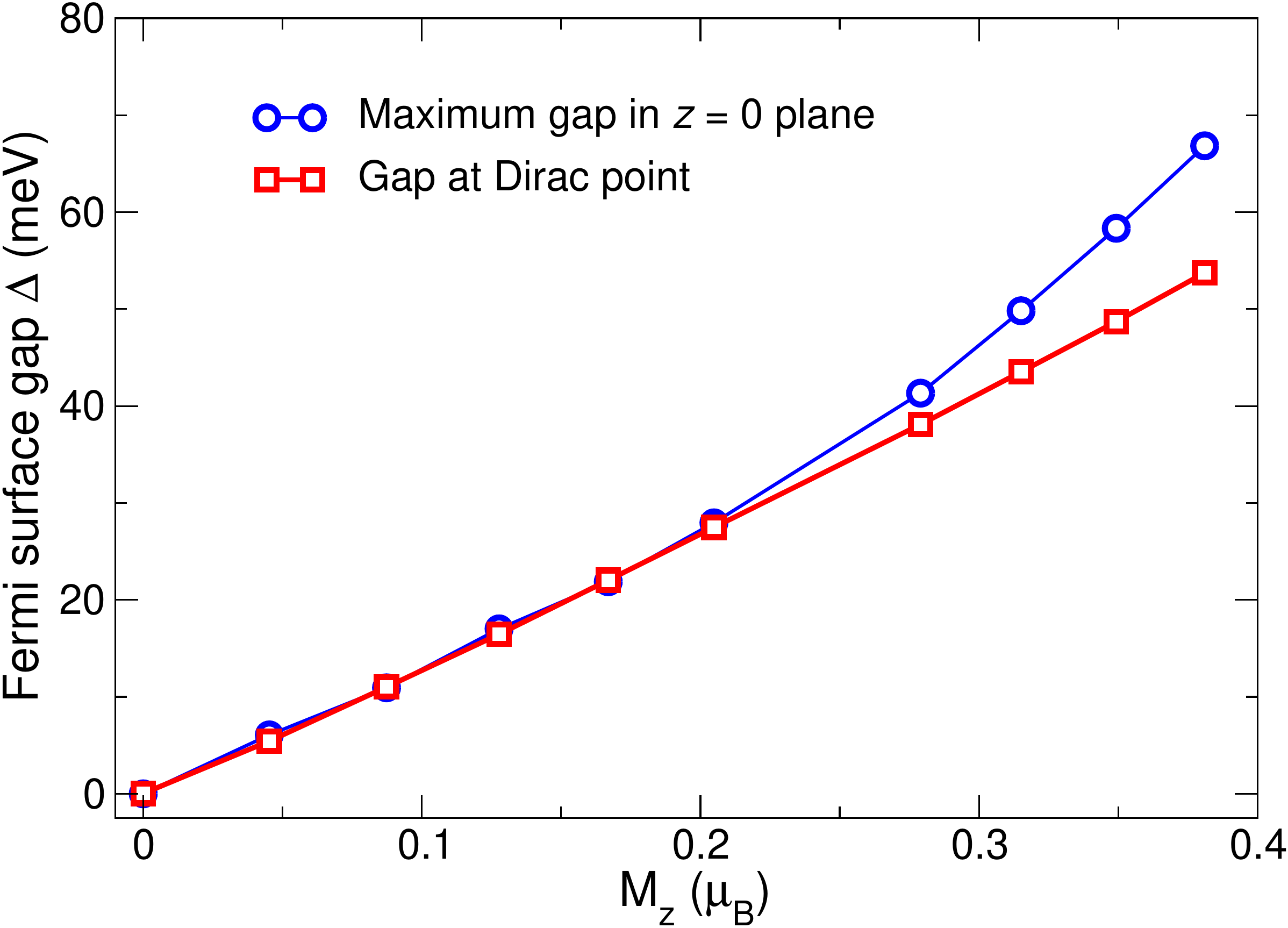}
  \caption{(Color online) Size of the  FS gap of {\urusi} in the $z =0$ plane, computed as a function of the amplitude of the static AF uranium moment along the $c$ axis.
  \label{fig:Gap}}
\end{figure}

Although the existence of the intense AF mode in the HO phase is evident, its connection to the HO phase is not yet fully understood.
A major question in the discussion is whether a mode of long-lived AF fluctuations can induce any  FS gapping.
In ordinary materials, a mode of (normally {\it transversal}) spin excitations would be a very weak perturbation of the electronic structure, and hence  it cannot produce any significant transition in the bulk thermodynamic or transport properties. However, {\urusi} is an exceptional material, as we find that a substantial FS gapping of 750~K occurs through symmetry breaking on an energy scale of less than 7~K.
To investigate the influence of the inelastic mode we consider the fluctuating longitudinal magnetic moment on one of the two uranium sublattices, i.e., 
%that are connected by $Q_{AF}$, i.e., 
$M_z (t) \approx M_0 \cdot {\rm cos} (\omega t)$.  $M_0$ is the maximal amplitude of the longitudinal  mode which we assume to be the same as the static LMAF moment of 0.39 $\mu_B$.
We furthermore assume for simplicity a dispersionless  frequency $\omega (q)\approx \omega $.
This is a good approximation near $\boldsymbol{Q}_{AF}$. 
Obviously, the resulting average sublattice moment, i.e., averaged over a sufficiently long time-scale $\tau$, 
${\overline M_z} = \tau^{-1} \int M_z (t) dt$, will vanish.

Next we consider what happens to the FS hot spots due to the presence of the longitudinal mode. At discrete time steps we can treat the AF mode as a frozen static antiferromagnetic arrangement with a certain amplitude of the moment $M_z$.  At each time snap-shot, the static AF arrangement will affect the FS hot spots, such that the degeneracy of bands at the Dirac point will be lifted. As the degeneracy at the Dirac point is quite sensitive, already for small static AF moments a FS gapping appears. In Fig.\ \ref{fig:Gap} we plot the size of the FS gap as a function of the static AF moment. The gapping appears for small AF moments only at the FS hot spots, but for increasing moments the gap widens until the hole FS part centered along the $\Gamma -$M direction has been gapped (cf.\ Fig.\ \ref{fig:nesting-AFM}). The magnitude of the computed gap develops, to a reasonable approximation, as a linear function of $M_z$.\cite{elgazzar09}
%the $z$-component of the moment.\cite{elgazzar09}
 Fig.\ \ref{fig:Gap} additionally shows that the maximal FS gapping shifts from being at the hot spots for moderate $M_z$ amplitudes to being along the $\Gamma -$M axis for the largest amplitude.
For the time-dependence of the gap caused by moderately slow spin-fluctuations it is essential to note that the gap $\Delta$ is an even function of $M_z$. Using that the magnitude of the calculated  gap is a linear function of the moment $M_z$,
% (Ref.\ \onlinecite{elgazzar09}),
 we can write $\Delta (t) = \alpha | M_z (t) |$, with $\alpha$ a positive constant. The time-dependence of the moment gives the time-dependence of the gap, $\Delta (t) =
 \Delta_0 | {\rm cos} ( \omega t) |$, with $\Delta_0 = \Delta_{_{\rm LMAF}}$, the maximal gap occurring for the LMAF phase.
 The time-averaged, macroscopic gap ${\overline \Delta}_{_{\rm HO}}$ follows from
 \begin{equation}
 {\overline \Delta}_{_{\rm HO}} = \frac{1}{\tau} \int \Delta (t) dt = \frac{\Delta_{0}}{\tau} \int | {\rm cos} (\omega t)| dt 
 = \frac{2}{\pi} \Delta_{_{\rm LMAF}}.
 \end{equation}
 Hence, with these approximations we estimate that the time-averaged HO gap is 64\% of the gap of the LMAF phase. This value is in reasonable agreement with experimental data obtained from resistivity measurements that find that the HO gap is  70 to 80\% of that of the LMAF phase.\cite{mcelfresh87,jeffries07,jeffries08,motoyama08}

 Consequently, in the special case of {\urusi} a long-lived AF mode can indeed induce a substantial FS gapping.
 Moreover, the similarity detected experimentally in the FS gapping of the HO and LMAF phases can be explained by the presence of the AF mode in the HO phase and AF order in the LMAF phase.
 %, which induces a substantial FS gapping,
% $\Delta_{_{\rm HO}}$. 
 Nonetheless, in spite of the similar gapping, thermodynamic and transport properties, we emphasize that the HO and LMAF are distinct phases. In the LMAF phase the sublattice U moment  $M_0$ is non-zero and thus is the standard order parameter (OP) for an ordered AF material.  In the HO phase, conversely, ${\overline M_z} =0$ and can obviously not be a suitable order parameter for the HO.  We note that in the here-developed theory it is the amplitude $M_0 (T)$ of the longitudinal AF mode that determines the magnitude
  of the time-varying gap (cf.\ Fig.\ \ref{fig:Gap}). An appropriate, observable secondary OP for the HO that follows from our model could be derived from the macroscopic average of an even function of  $M_z (t)$. 
  As mentioned above, the FS gap is such an even function, $\Delta (t) \propto | M_z (t) |$.
  Assuming for  the temperature-dependence of the mode 
$M_z (t) = M_0 (T) \cdot {\rm cos} (\omega t)$, this gives for the temperature-dependence of the averaged, macroscopic gap ${\overline \Delta}_{\rm HO} (T) \propto M_0 (T)$.
Consequently,  if the amplitude of the mode $M_0 (T)$ behaves as an OP, the macroscopic, time-averaged gap  ${\overline \Delta}_{\rm HO} $ should be expected to behave as a secondary OP. Two experimental techniques, point contact and far-infrared optical spectroscopy revealed that the HO gap does indeed approximately behave as a typical BCS-type order parameter with temperature.\cite{hasselbach92,escudero94,thieme95} Very recent STS measurements have unambiguously demonstrated this feature.\cite{schmidt10,aynajian10} We remark that in contrast to standard BCS theory the here-obtained FS gap is not symmetric around $E_F$, see Fig.\ \ref{fig:bands-123}.

A second possibility for an even function of $M_z (t)$ would be the dynamical susceptibility, expressed by the (longitudinal) dynamical spin-spin correlation function,
\begin{equation}
S ({\mbox{\boldmath$q$}}, \omega' ) \propto \sum_{i,j} e^{i{\mbox{\boldmath$q$}} \cdot 
({\mbox{\boldmath$R$}}_i - {\mbox{\boldmath$R$}}_j )}
\int e^{-i\omega' t}
\langle S_{z,i}(t) S_{z,j} (0) \rangle
% M_{z,i} (t) M_{z,j} (0) 
dt,
\label{eq:spin-spin}
\end{equation}
where $S_{z,i}$ is the $z$-component of the spin operator at the U position ${\mbox{\boldmath$R$}}_i$. For relativistic materials such as actinides, it is more appropriate to use the total angular momentum, i.e.,
$\langle J_{z,i}(t) J_{z,j} (0) \rangle$. Approximating  $\langle J_{z,i}(t) J_{z,j} (0) \rangle \approx $ 
$\langle J_{z,i}(t)\rangle \langle J_{z,j} (0) \rangle$, i.e., as $M_{z,i} (t) M_{z,j} (0)$, and using for
${\mbox{\boldmath$q$}}$ the AF wavevector
${\mbox{\boldmath$Q$}_{AF}}$, the dynamical spin-spin correlation function $S ({\mbox{\boldmath$q$}}, \omega' )$ would show a resonance peak at $\omega ' = \omega$
that would be detectable in inelastic neutron experiments. 
Computing the frequency-integrated area of the peak, $A = \int S ( {\mbox{\boldmath$Q$}_{AF}}, \omega' ) d \omega '$, it follows that  $A (T) \propto  M_0^2 (T) \propto {\overline \Delta}^2 (T)$.
Hence, the area of the inelastic neutron peak at the AF wavevector was predicted to behave as a secondary OP.\cite{elgazzar09}  
%Encouraging results for such an OP behavior were obtained recently from inelastic neutron experiments.\cite{villaume08,wiebe07} 
A very recent, precise study of the inelastic AF resonance confirms that the area of the inelastic peak indeed behaves as a BCS-type OP.\cite{bourdarot10}
In the LMAF phase the AF mode seizes to exist and hence the inelastic peak area vanishes. Adopting the spin-spin correlation function expressed through the inelastic peak area $A$ as a derived OP for the HO phase, one obtains the situation where the staggered AF moment $M_z \neq 0$ and $A=0$ in the LMAF phase, while conversely $M_z = 0$ and $A \neq 0$ in the HO phase, i.e., the phases have distinct OP's (cf.\  the discussion regarding this in Refs.\ \onlinecite{shah00}, \onlinecite{niklowitz10}, \onlinecite{motoyama03}, \onlinecite{bourdarot03}, \onlinecite{bourdarot05}, and \onlinecite{uemura05}).

Interestingly, so far only the gap $\Delta_{_{\rm HO}}$ and the inelastic neutron peak intensity $A$ were proven to display typical OP behavior.\cite{bourdarot10,aynajian10} Both quantities are evidently related to the intense AF mode in the HO phase, emphasizing that the AF mode is essential for the HO.
Earlier $^{29}$Si NMR experiments\cite{bernal01} indicated that the NMR line width displayed OP behavior in the HO phase. Another NMR experiment, however, observed an inhomogeneous development of AF order below $T_0$,\cite{matsuda01} and a further NMR experiment did indeed detect an increase of the line width below $T_0$, but this effect disappeared at lower temperatures and more strain-free crystals.\cite{takagi07} 

Second order phase transitions are characterized by a divergence of a (generalized) susceptibility at the critical temperature. 
 We note that the critical behavior of the dynamical susceptibility $\chi (\boldsymbol{q}, \omega)$ at the HO transition is consistent with the proposed importance of spin-fluctuations.
Its imaginary part $\chi^{\prime\prime} $ is proportional to the spin-spin correlation function, 
$S(\boldsymbol{q}, \omega)$. Its real part $\chi^{\prime} (\boldsymbol{q}, \omega)$, however, was found--from inelastic neutron experiments--to exhibit a sharp cusp at $\boldsymbol{q}=\boldsymbol{Q}_{AF}$ and $\omega \approx 0$ when entering the HO.\cite{mason95} This signals a divergence of  $\chi^{\prime} (\boldsymbol{Q}_{AF}, \omega \approx 0)$ at $T_0$, which has been broadened only by experimental resolution.

The magnetic fluctuations could also bear relevance\cite{elgazzar09} to the occurrence of unconventional\cite{matsuda96,kasahara07,yano08} SC out of the HO at temperatures below 1.2~K.\cite{palstra85,maple86,schlabitz86}
The persistence of spin-fluctuations at the incommensurate wavevector down into the SC phase was observed by Broholm {\it et al.}\cite{broholm87,broholm91} A very recent inelastic neutron study\cite{hassinger10} reveals a small energy shift in the AF mode upon entering the SC phase, suggesting an involvement of the AF spin excitations in mediating the SC.

\subsubsection{Symmetry breaking}

%OP differences and symmetry breaking
Static AF ordering obviously corresponds to breaking of time-reversal symmetry and of the body-centered translation vector in the LMAF phase; the latter causes a doubling of the unit cell along the $c$ axis.
What would be the symmetry breaking induced by the longitudinal AF mode?  The unconventional answer is, that it in fact depends on the time-scale being considered. On a very short time-scale, typically much less than that of one oscillation, the body-centering as well as time-invariance will ``instantaneously" be broken. On a sufficiently long time-scale however the moments on both uranium atoms have reverted their orientation many times, and thus the two atoms are no longer distinguishable.  As the two U atoms are undistinguishable on long time-scales there will, consequently, be no breaking of time-reversal symmetry nor of the $c$-axis  translation vector. Alternatively, this can be assessed, for example, from the moment-moment correlation function (Eq.\ (\ref{eq:spin-spin})), which is time-even. Most experimental techniques probe quantities on relatively long time-scales, much longer than picoseconds (e.g., $\mu$SR or elastic neutrons). These will not be able to detect a symmetry breaking in the HO phase.
% in comparison to the PM phase.  

The question whether time-reversal or possibly translational symmetry is broken in the HO is an issue of on-going debate. 
Earlier neutron experiments provided evidence for the breaking of time-reversal symmetry,\cite{walker93}
however these measurements might have been influenced by the presence of parasitic small moments. Also later, more precise neutron experiments deduced evidence for time-invariance breaking,\cite{bourdarot03,bourdarot05} yet the issue is not unambiguously settled.
Newer $^{29}$Si NMR on single crystals detected a small line shift as well as a linewidth enhancement below $T_0$ which corresponds to the presence of a small  internal magnetic field.\cite{bernal06} This finding might indicate the presence of time-invariance breaking in the HO phase.
Also, the temperature-dependent static magnetic susceptibility of {\urusi} displays a clear kink at $T_0$.\cite{park97,pfleiderer06,ramirez92} Such feature is not uncommon at an AF ordering transition in actinide compounds, and it indicates the influence of a magnetic field on the HO.  This in turn could suggest that a purely electric OP such as an electric quadrupole or hexadecapole is unlikely.
Electric hexadecapolar order has been proposed for PrRu$_4$P$_{12}$ precisely because of the observed insensitivity of an ordering transition to a magnetic field.\cite{takimoto06}

Altogether, clear experimental evidence for either breaking of the time-reversal or the translational symmetry in the HO phase is so far lacking.  
A possible verification of the symmetry-breaking due to the AF mode would require a sufficiently fast experimental technique, such as x-ray scattering or photoemission. 
The here-proposed breaking of the body-centering could be probed in ARPES experiments, where the corresponding doubling of the unit cell would appear as a sudden folding of bands at $T_0$, together with a change in the $c^{\star}$ axis periodicity.
A further argument in favor of time-reversal and translational symmetry breaking in the HO might come from dHvA and SdH experiments. The theoretical FS (Fig.\ \ref{fig:FS}) computed for the symmetry-broken phase agrees well with the inferred experimental FS. This correspondence  suggests that the body-centered translation vector is broken in the HO phase and most likely the time-reversal symmetry as well. Conversely, using the FS computed for the \textsc{bct} phase, it is not possible to explain the observed dHvA and SdH quantum oscillations.\cite{ohkuni99,jo07,shishido09} Although this is consistent with the dynamical symmetry breaking model, it cannot be excluded that other mechanisms for the HO lead to the same symmetry reduction.

\section{Discussion}

It is elucidative to consider if hidden order phases have been discovered in other $5f$ materials.
 %In only a few $5f$ electron materials have hidden ordered phases been discovered. 
 Apart from {\urusi}, the actinide oxide NpO$_2$ has drawn considerable attention,\cite{paixao02,kubo05,santini06,santini09} because a sharp phase transition to an unusual phase occurs 
below 30~K as is witnessed by specific heat measurements.\cite{osborne53} Intensive  research during the last two decades revealed that, in the absence of  any magnetic dipole moment, long-range multipolar ordering of a higher-order magnetic multipole (octupole or higher) on the Np ions is likely to occur in the unusual phase below 30~K.\cite{paixao02,santini09} These higher multipoles cannot be observed experimentally, but electric quadrupolar order appearing as a {\textit{secondary}} OP has been observed through resonant x-ray scattering (RXS)\cite{paixao02} and $^{17}$O NMR.\cite{tokunaga05}
Theories based on $5f^3$ localized states on the Np$^{4+}$ ion have further provided insight in the multipolar order.\cite{santini09,kuramoto09}

Also for {\urusi} a number of theories for the low-temperature order have been developed on the basis of localized or nearly localized $5f^2$ configurations.
Specifically, for {\urusi} the following multipolar OPs have been proposed: electric quadrupole,\cite{santini94,ohkawa99,harima10} magnetic octupole,\cite{kiss05,fazekas05,hanzawa05,hanzawa07} electric hexadecapole,\cite{haule09}  and magnetic triakontadipole.\cite{cricchio09}
Other theories on the  basis of crystal electrical field (CEF) considerations for a U$^{4+}$ $5f^2$ ionic state have also been developed.\cite{sikkema96,okuno98}
Electric quadrupolar ordering should be detectable with RXS,\cite{paixao02} yet experimental studies gave a null result.\cite{amitsuka10,caciuffo-unp} Quadrupolar order requires an $E_1 - E_1$ scattering channel to be detected, whereas higher order multipoles require at least an $E_2$ optical transition. Since the scattering cross-section for $E_2$ transitions is extremely small, a detection of higher order multipoles is currently unlikely.
Furthermore, most of these theories are based on a localized U $5f^2$ configuration, but in spite of a similarity in the thermodynamic properties of {\urusi} and NpO$_2$, treating the $5f$'s as localized in NpO$_2$ is a good approximation,\cite{suzuki10} but
%, in the light of the above presented calculated results, 
it is questionable if this approximation is valid for {\urusi}, too. 

Triakontadipolar order in the HO phase of {\urusi} was recently proposed\cite{cricchio09}
on the basis of LSDA+$U$ calculations with a Coulomb $U$ of about 1 eV. In contrast to localized $5f^2$ theories, the $f$ electrons are relatively delocalized in this description.
These LSDA+$U$ calculations predict an unusual long-range ordered AF state for {\urusi} with {\it parallel} spin and orbital magnetic moments.\cite{cricchio09b}  {\urusi} is an unconventional material and it might indeed be that such an unusual magnetic state is realized. It should be noted however that thus far not a single magnetically ordered U-compound has been discovered having this property. This can be understood from the strong spin-orbit interaction in actinides, which enforces always antiparallel spin and orbital moments in the early actinides. Also, conventional LSDA calculations predict antiparallel spin and orbital moments,\cite{elgazzar09} and, as mentioned above, antiparallel spin and orbital moments are in fact consistent with neutron form factors measurements.\cite{kuwahara06}
Moreover, the values deduced for the separate spin and orbital moments are in close agreement with values predicted by LSDA calculations. The Fermi surface predicted by LSDA+$U$ calculations deviates already substantial from the LSDA FS. In particular, new band crossings appear and the FS gapping feature is lost  (cf.\ Fig.\ \ref{fig:bands+U}); thus, the LSDA+$U$  approach with a $U$ of more than 1 eV cannot explain the FS gapping appearing in the LMAF phase. The recent LSDA+$U$ calculations\cite{cricchio09} predict that, with expansion of the lattice parameter $a$, a phase transition would occur in {\urusi} from the AF ordered phase to the same AF ordered phase, but with both spin and orbital moments being opposite to the original ones. The triakontadipolar moment is predicted\cite{cricchio09} to be present in both AF phases, but it could become the main OP precisely at the transition point between these two AF phases, where the dipolar moment would vanish. 
It should be noted, however, that such a phase transition from one AF to another AF phase, being identical by symmetry has not been observed in {\urusi}.  Also the predicted\cite{cricchio09,cricchio09b} parallel spin and orbital moments are not supported by experimental observations.

Electric hexadecapolar order in {\urusi} has recently been proposed on the basis of DMFT calculations.\cite{haule09} These calculations suggest a nearly localized $5f^2$ configuration on the uranium ion, consisting of two singlet CEF levels, separated by 35~K. The HO phase emerges from an excitonic mixing of the CEF ground and the first excited state singlet. The same CEF level scheme (but notably with ground and excited state reversed) had been proposed\cite{nieuwenhuys87} earlier to explain the Ising-like anisotropy and maximum of the magnetic susceptibility. Although such CEF model can explain certain properties of {\urusi}, it is one of the observations of the present study that the CEF model does not corroborate with many other experimental data. 

The here-reported computed results suggest altogether that an applicable explanation of the properties of {\urusi} ought to arise from an itinerant $5f$ electronic structure. 
A number of earlier HO theories\cite{barzykin95,ikeda98,chandra02,mineev05,varma06,kotetes-unp} have been based on the itinerant $5f$ picture, and more recently, two theories have emphasized the importance of dynamic spin-fluctuations for the HO.\cite{elgazzar09,balatsky09}
The earlier theories\cite{barzykin95,ikeda98,chandra02,mineev05,varma06} could offer an explanation for some aspects of the HO, but none of the models could be unambiguously confirmed experimentally (see, e.g., Refs.\ \onlinecite{villaume08}, \onlinecite{biasini09}, and \onlinecite{wiebe04} for a discussion). 
The two HO theories that highlight the role of dynamical spin excitations have either focussed on spin-fluctuations at the AF wavevector\cite{elgazzar09} or  at the incommensurate wavevector.\cite{balatsky09}
Obviously, spin-fluctuations at the incommensurate wavevector as proposed in the recent theory by Balatsky {\it et al}.\cite{balatsky09} are expected to have an influence, therefore further investigations of the incommensurate mode are needed to establish the relative importance of the two modes. 

In the present work we have studied particularly the effect of the AF mode and find that its presence in the HO phase can explain the FS gapping, the broken-symmetry FS, the relation $\Delta_{_{\rm HO}} \approx (0.7-0.8) \Delta_{_{\rm LMAF}}$,
as well as the mean-field OP behavior of $\Delta_{_{\rm HO}}$ and of the inelastic neutron peak.
% between the gaps in the HO and LMAF phases. 
The dynamical symmetry breaking model nonetheless builds on the surprising and exceptional explanation of the HO phase being driven by the dynamic AF mode. Spin excitations are usually only weak perturbations of the ground state, that in conventional materials, cannot modify the ground state nor its properties. 
In  {\urusi}, however, the exceptional situation appears to be realized that the low-lying spin excitations actually dictate the thermodynamic properties of the HO phase. Thus far a similar situation was apparently realized only in one other material, PrAu$_2$Si$_2$,\cite{goremychkin08} which has the same crystal structure as {\urusi}. 
Why the low-lying AF spin excitations are so effective in {\urusi} is related to materials' specific aspects. The calculated energy scale of AF excitations is only of the order of 7~K, but they couple to an unexpectedly large FS reconstruction, with gaps of about 700~K. Thereby these low-lying modes can essentially modify the thermodynamic and transport properties and are indeed inherent to the HO phase.

%In addition to supporting the itinerant $5f$ picture, 
Our calculations lastly emphasize that in {\urusi} several remarkable, {\textit{materials' specific}} features are combined. First, the LMAF phase with its relatively modest total moment has a total energy that is very close to that of the PM phase.
% (only 7~K per formula unit deeper). 
Second, the low-temperature FS of {\urusi} supports a nesting vector $\boldsymbol{Q}_{AF}$ that promotes long-range AF ordering or longitudinal AF spin-fluctuations. Third, the energy bands in the {\textsc{st}} BZ exhibit accidental degenerate crossing points at $E_F$ as well as close to $E_F$. Fourth, a breaking of the $c$ axis translational symmetry together with time-reversal symmetry breaking in the LMAF phase or through longitudinal spin-fluctuations causes a lifting of the degeneracy at the Dirac points and thus to the opening of a partial gap. 
%Longitudinal AF spin-fluctuations may cause the opening of gap in the HO. 
These observations underline that to a large extent the HO is not a generic property, but rather, it is borne out of the materials' specific electronic structure.

\section{Conclusions}

For a quarter century the mysterious hidden order phase in {\urusi} has been in the focus of many investigations. A detailed understanding of the underlying electronic structure of {\urusi} in both the PM and LMAF phases is required before a full explanation of the HO can be formulated. Neither the PM nor the LMAF phase appear to be exceptional, therefore first-principles electronic structure calculations should be capable of explaining the known solid-state properties of these two phases. Using {\it ab initio} electronic structure calculations on the basis of the DFT-LSDA/GGA, the GGA+$U$, and LDA+DMFT methodologies, we have performed an extensive study of the electronic structure of {\urusi}.
A major question has for a long-time been what the applicable electronic structure of the PM and LMAF phases is, whether it has itinerant or localized uranium $5f$ electrons or perhaps $5f$ electrons with dual, i.e., both itinerant and localized characteristics 
(cf.\ Ref.\ \onlinecite{petit09}).
Our conclusion regarding this issue is that we obtain an electronic structure picture consistent with available experimental data only when we adopt delocalized $5f$ states.

Specifically, we find that materials' specific DFT-LSDA/GGA electronic structure calculations on the basis of delocalized $5f$'s explains the following low-temperature properties of {\urusi}:
1) the equilibrium volume, 2) the internal $z_{\rm Si}$ coordinate and $c/a$ ratio,
3) the bulk modulus and equation of state, 4) the spin and orbital magnetic moment of the LMAF phase, 5) the closeness in total energy of the PM and LMAF phases, with the LMAF phase becoming more stable under pressure,  6) the Fermi surface gapping and instability, 7) the compensated metal character, 8) the number of holes, 9) the $5f$ occupation number, 10) the resistivity jump at the PM to LMAF transition, 11) crystallographic anisotropy of the resistivity, 12) the gapping in the infrared optical spectra, 13) the antiferromagnetic and incommensurate FS nesting vectors, and 14) the dHvA and SdH frequencies, as well as their angular dependence.
In contrast, assuming a localized uranium $5f^2$ configuration we cannot obtain a satisfactory explanation of the experimental low-temperature data.  
The recent ARPES data\cite{santander09} form a notable exception to the set of low-temperature properties that are explained by $5f$ itinerant electronic structure calculations. A (nearly) localized $5f^2$ uranium configuration appears to tally better with the observed spectral features.\cite{haule09} As the Fermi surface detected through ARPES  evidently has to match that obtained with quantum oscillation techniques, more experimental investigations appear to be needed to resolve this issue.

We have additionally studied {\urusi} in the high-temperature PM phase, using LDA+DMFT calculations. Our DMFT calculations predict the progressive opening of a quasi-particle coherence gap at the chemical potential when temperature is reduced toward the coherence temperature.

An explanation of the hidden order requires still taking a step beyond conventional electronic structure calculations, however, the electronic structure picture underlying any explanation of the the HO phase must be consistent with those underlying the PM and LMAF phases. The current investigations strongly emphasize that a well-grounded explanation of the HO has to emerge from an itinerant $f$-electron picture.
From the presented calculations we  conclude that the presence of the intense  AF mode in the HO phase could explain ($i$) the FS gapping occurring in the HO phase, ($ii$) its broken-symmetry FS extremal orbits and incommensurate nesting vector, ($iii$) the experimentally found relation between the gaps in the HO and LMAF phases, $\Delta_{_{\rm HO}} \approx (0.7-0.8) \Delta_{_{\rm LMAF}}$, and $(iv)$ the OP behavior of the gap $\Delta_{_{\rm HO}}$ and the inelastic dynamical susceptibility.
The dynamical symmetry breaking model appears so far to be the only theory that explains and even predicted\cite{elgazzar09} these four properties.
%relation $\Delta_{_{\rm HO}} \approx (0.7 - 0.8)\Delta_{_{\rm LMAF}}$.
The importance of this mode for bringing about the HO transition is further exemplified by its occurrence and gapping in its magnon dispersion, its assistance in the entropy removal, as well as the divergent behavior of the staggered susceptibility $\chi^{\prime}$
% (\boldsymbol{q}, \omega \approx 0)$ 
at $T_0$.
The presence of the inelastic AF mode at low-temperatures, down into the SC phase,\cite{hassinger10} could provide a clue as to why an unconventional form of superconductivity develops.

\begin{acknowledgments}
During the course of this work we have benefited from discussions with J.\ W.\ Allen, H.\ Amitsuka, A.\ Balatsky, N.\ Bernhoeft, M.\ Biasini, F.\ Bourdarot, W.\ J.\ L.\ Buyers, R.\ Caciuffo, P.\ Chandra, P.\ Coleman, N.\ J.\ Curro, J.\ D.\ Denlinger, 
%T.\ Durakiewicz, 
J.\ Flouquet, M. Graf, H.\ Harima, V.\ Janis, J.\ R.\ Jeffries,  G.\ H.\ Lander,  N.\ Magnani, M.\ B.\ Maple, Y.\ Matsuda, K.\ McEwen, Y.\ Onuki, R.\ Osborn, A. Santander-Syro, and J.\ Schoenes. We also gratefully acknowledge a discussion on Fermi surface orbits with E.\ Hassinger and G.\ Knebel.
This work has been support through the Swedish Reserach Council (VR), STINT, EU-JRC ITU, and the Swedish National Infrastructure for Computing (SNIC).
 \end{acknowledgments}


\begin{thebibliography}{99}
\bibitem{note1}
Note that we have included the additional points F, $\Sigma$, and Y that strictly do not belong to the standard \textsc{bct} high-symmetry points.
\bibitem{disclaimer}
The computed $5f$ occupation number is obtained for overlapping atomic spheres. Non-overlapping muffin-tin spheres give a somewhat smaller $5f$ number in the spheres.
\bibitem{note2}
The large FS computed for the $5f$ localized configuration is in itself an interesting feature, because conventional understanding associates localized $f$ states with a small FS, while delocalized $f$'s are associated with a large FS. We do not find such standard behavior for {\urusi} on account of the band re-hybridization which removes a large FS part below $T_0$ and $T_N$.
\end{thebibliography}
\end{document}